\documentclass[namedreferences]{solarphysics}
\usepackage[optionalrh,solaromanenum]{spr-sola-addons} % For Solar Physics 
\usepackage{solaheader}
\usepackage{graphicx}                    % For eps figures, newer & more powerfull
\usepackage{amssymb}                    % useful mathematical symbols
\usepackage{color}                       % For color text: \color command
\usepackage{caption}
\usepackage[rightcaption]{sidecap}
\sidecaptionvpos{figure}{t}
\usepackage[normalem]{ulem}
\usepackage{hyperref}
\usepackage{breakurl}

\newcommand{\deriv}[2]{\frac{{\mathrm d} #1}{{\mathrm d} #2}}
\newcommand{\pder}[2]{\frac{\partial #1}{\partial #2}}

\newcommand{\solareceiveddate}{11 December 2014}
\newcommand{\solaaccepteddate}{21 September 2015}
\newcommand{\solaonlinedate}{13 November 2015}
\newcommand{\Xdoi}{s11207-015-0792-y}

\begin{document}

\begin{article}

\begin{opening}

\title{Improved Helioseismic Analysis of Medium-$\ell$ Data from the \textit{Michelson Doppler Imager}}

%%%%%%%%%%%%%%%%%%%%%%%%%%%%%%%%%%%%%%%%%%%%%%%%%%%
%% Authors Names
%
\author{Timothy P.~\surname{Larson}$^{1}$\sep
        Jesper~\surname{Schou}$^{2}$ 
       }

%%%%%%%%%%%%%%%%%%%%%%%%%%%%%%%%%%%%%%%%%%%%%%%%%%%
%% Runningheads
%
\runningauthor{T.P. Larson and J. Schou}
\runningtitle{Improved Medium-$\ell$ Analysis}

%%%%%%%%%%%%%%%%%%%%%%%%%%%%%%%%%%%%%%%%%%%%%%%%%%%
%% Affilations 
%
  \institute{$^{1}$ Stanford University, Stanford, California, USA email: \url{tplarson@sun.stanford.edu}\\
             $^{2}$ Max-Planck-Institut f\"{u}r Sonnensystemforschung, G\"{o}ttingen, Germany
%                    ~~~email: \url{schou@mps.mpg.de}
             }

%%%%%%%%%%%%%%%%%%%%%%%%%%%%%%%%%%%%%%%%%%%%%%%%%%%
%%% Abstract 
\begin{abstract}
We present a comprehensive study of one method for measuring various parameters of 
global modes of oscillation of the Sun. Using velocity data taken by the 
{\it Michelson Doppler Imager} (MDI), we analyze spherical harmonic degrees $\ell \leq 300$. 
Both current and historical methodologies are explained, and the 
various differences between the two are investigated to determine their effects on global-mode 
parameters and systematic errors in the analysis.  These differences include a number of 
geometric corrections made during spherical harmonic decomposition; updated routines for generating 
window functions, detrending timeseries, and filling gaps; and consideration of physical effects such 
as mode profile asymmetry, horizontal displacement at the solar surface, and distortion of eigenfunctions by 
differential rotation.  We apply these changes one by one to three years of data, and then reanalyze 
the entire MDI mission applying all of them, using both the original 72-day long timeseries and 360-day long
timeseries. We find significant changes in mode parameters, both as a result of the various changes 
to the processing, as well as between the 72-day and 360-day analyses.  We find reduced residuals 
of inversions for internal rotation, but seeming artifacts remain, such as the peak in the 
rotation rate near the surface at high latitudes. 
An annual periodicity in the $f$-mode frequencies is also investigated.
\end{abstract}

%%%%%%%%%%%%%%%%%%%%%%%%%%%%%%%%%%%%%%%%%%%%%%%%%%%
%% Keywords
%

\end{opening}
%-------------------------------------------------

%%%%%%%%%%%%%%%%%%%%%%%%%%%%%%%%%%%%%%%%%%%%%%%%%%%
%% Sections
%
% \section{}%\label{s:?} 
\section{Introduction}
     \label{S-intro} 

The {\it Michelson Doppler Imager} (MDI: \opencite{scherrer95}) onboard the {\it Solar and Heliospheric
Observatory} (SOHO) took data from December 1995 to April 2011.  Equipped
with a $1024 \times 1024$ CCD, it was capable (in full-disk mode) of sending
down dopplergrams with a spatial resolution of 2.0 arcsec per pixel at a
cadence of 60 seconds using the Ni\,{\sc i} 6768 \AA \ spectral line.  However, due to
telemetry constraints, MDI was operated in full-disk mode for only a few
months total each year.  For the rest of the time, we have only data that
were convolved in each direction onboard the spacecraft with a 
gaussian vector of 21 integers,  subsampled by a
factor of five, and cropped to 90\,\% of the average image radius
in order to fit into the available telemetry
bandwidth.  It was these dopplergrams that comprised the
Medium-$\ell$ Program and acquired the label of \textsf{vw\_V} for ``vector-weighted 
velocity''.  The \textsf{vw\_V} data were the input to all of the analysis described
here.  For overviews of global mode helioseismology, the reader is referred 
to \inlinecite{jcd2004} and \inlinecite{gough2013}.

Dopplergrams are decomposed into spherical harmonic components
described by their degree [$\ell$] and azimuthal order [$m$], which are formed
into timeseries and Fourier transformed.  We work in the medium-$\ell$ regime, 
which is defined as the range where peaks
in the power spectrum, corresponding to the oscillation modes,
 are well-separated from those of different degrees. 
Sets of modes with
the same radial order [$n$] form ridges; modes with $n=0$ are labelled
$f$-modes, and those with $n>0$ are labelled $p$-modes.
The medium-$\ell$ regime is conventionally
taken to be $\ell \leq 300$ for the $f$-modes and $\ell \leq 200$ for the $p$-modes. 
The Fourier transforms are fit
to yield the mode frequencies (among other parameters) for multiplets
described by $\ell$ and $n$. 
In a spherically symmetric Sun, the 
frequency would be the same for all $m$.  Asphericities such as rotation lift this degeneracy, 
and the variation of frequency with $m$ can be fit by a polynomial, resulting 
in the so-called $a$-coefficients
(see Section \ref{S-pkbgn}). The
frequencies and $a$-coefficients can be inverted to infer the sound speed
or angular velocity in the solar interior as a function of latitude and
radius.  In this work we have used the odd $a$-coefficients to perform
regularized least squares (RLS) inversions for angular velocity. 
The RLS method attempts to balance fitting the data with the smoothness of the solution,
since the inverse problem is ill-posed \cite{schou94inv}.

\begin{figure}
%\centerline{\includegraphics[width=0.8\textwidth,clip=]{a3bump2.eps}}
\centerline{\includegraphics[width=0.9\textwidth,clip=]{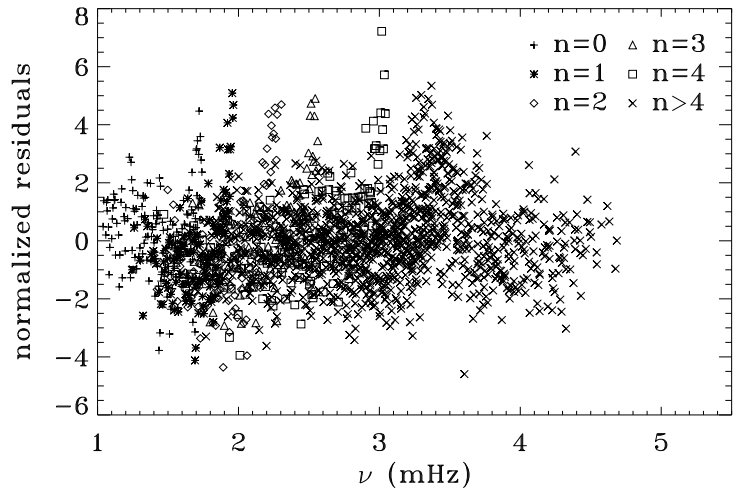}}

\caption{Residuals of $a_3$ coefficients normalized by their errors as a function of frequency for the
72-day interval beginning on 8 January 2004.
}

\label{F-bumporig}
\end{figure}

With an internal-rotation profile in hand, 
one can compute the corresponding $a$-coefficients.  
These inferred $a$-coefficients represent a 
fit to the measured $a$-coefficients.  We can use the residuals of this fit
to investigate potential systematic errors in the $a$-coefficients.
One problem with the original analysis can be seen in Figure~\ref{F-bumporig},
which presents the normalized residuals of $a_3$.  If the
model were a good fit to the data, one would expect these to be normally
distributed around zero with unit variance.  A significant deviation from
this expectation is the ``bump'' at around 3.4\,mHz, which can be seen in all of
the odd $a$-coefficients and their residuals, and alternates in sign between them.
%\footnote{The sign of the bump alternates between successive odd $a$-coefficients}.  
Furthermore, the shape of
the bump depends on the width of the frequency interval used in the mode fitting, which by
itself indicates a problem with the fits \cite{comparison1}. Also visible in this plot are
deviations from a continuous function at the ends of ridges.  This 
feature, known as ``horns'', is visible in several of the mode parameters
and is not reproducible by any reasonable internal-rotation profile (see
Section \ref{S-syserrs}). In the new analysis, the 
residuals have been substantially reduced, but the fact that they
are still quite large indicates that the errors are still dominated by
systematics (see Section \ref{S-syserrs}).

\begin{figure}
\centerline{\includegraphics[width=0.8\textwidth,clip=]{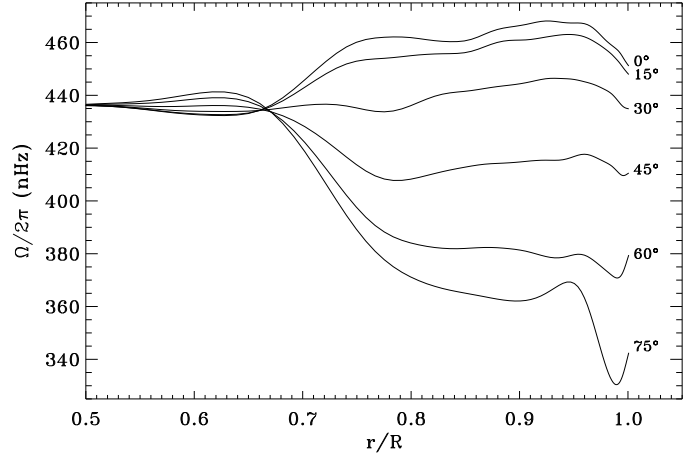}}

\caption{Rotation profile for the same time interval shown in Figure \ref{F-bumporig}.
The curves correspond to the latitudes indicated.
}
\label{F-rot2d}
\end{figure}

In parallel to the MDI analysis, the {\it Global Oscillation Network Group}
(GONG: \opencite{harvey96}) has done an independent medium-$\ell$ analysis of dopplergrams taken
from six ground-based observatories (the GONG network), using the same spectral line and cadence
as MDI.  Although the two analyses are generally in good agreement, in
certain areas the inferences drawn by the two projects differ by more than
their errors.  In particular, the above-mentioned bump is absent in the
GONG analysis. Likewise, the MDI analysis indicates a polar jet at a
latitude of about $75 ^{\circ}$, shown in Figure \ref{F-rot2d}, which is not seen in the GONG analysis.  
Excluding the modes that contribute to the bump does not remove this
high-latitude jet. Although the jet may be a real feature, the fact that
it is not seen in the full-disk analysis of MDI data makes this
questionable \cite{variations}. 
Until such discrepancies can be resolved, the analysis results
must remain in doubt, and the issue has been studied at length by several
investigators with little success \cite{comparison1}.

\begin{figure}
\centerline{\includegraphics[width=0.9\textwidth,clip=]{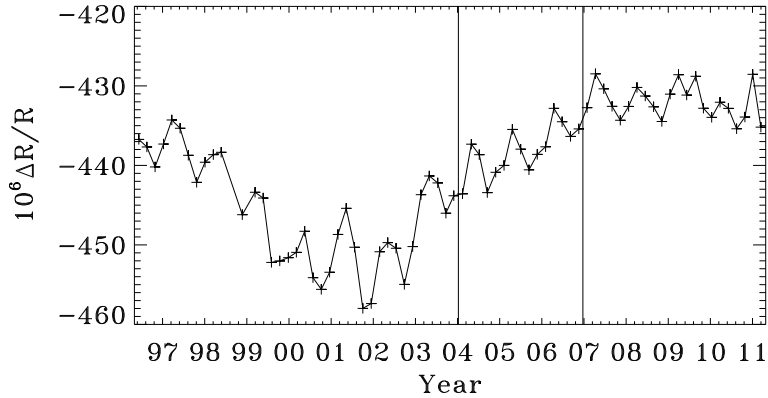}}

\caption{Fractional difference in seismic radius between observations and
a model as a function of time, averaged over all degrees $\ell$.  Vertical lines indicate the three years that 
we reanalyzed for each change in the processing (see Section \ref{S-Data}).}

\label{F-radius}
\end{figure}

Another apparent systematic error seen in the original MDI analysis is a
one-year periodicity in the fractional change in the seismic radius of the
Sun (see Figure~\ref{F-radius}), which is proportional to the fractional
change in $f$-mode frequencies \cite{antia01}.  This cannot be studied with
the GONG results because they do not fit enough $f$-modes, 
while the MDI full-disk data do not help either since they are only taken for approximately
one time interval (long enough for global analysis) per year.  Although it
was presumed that this effect had to do with an annual variation in
leakage (see Section \ref{S-method}) between the modes, early investigations revealed that using a
corrected $B_0$, $P_{\rm eff}$\footnote{The angle $P_{\rm eff}$ is the effective $P$-angle, which  
is the angle between the solar-rotation axis and the column direction
on the CCD; the angle $B_0$ is the heliographic latitude of the sub-observer point.}, 
and solar radius did not make a substantial difference \cite{reduction}.

It was to address all of these issues that a reanalysis of the medium-$\ell$
data was undertaken.  The original analysis was in general very
successful, but it is based on certain approximations.  Physical effects
such as mode profile asymmetry \cite{duvall93}, 
horizontal displacement of the near surface matter,
distortion of eigenfunctions by the differential rotation \cite{woodard89}, and a potential
error in the orientation of the Sun's rotation axis as given by the Carrington
elements \cite{beckgiles}, were not taken into account.  Likewise, instrumental effects
such as cubic distortion from the optics (see Section \ref{S-sht}), 
misalignment of the CCD with the solar rotation axis, 
an alleged CCD tilt with respect to the focal plane, and image-scale 
errors were ignored \cite{korzennik04}.  Furthermore, new algorithms for generating the
window functions, detrending the timeseries, and filling the gaps had
become available.  We updated the data analysis to include each of these
considerations in turn to see what effect, if any, they had on the mode
parameters and systematic errors.

In the next section we describe the datasets that we analyzed and how.  In Section \ref{S-method} 
we give a detailed description of all of the steps in the data analysis.  Section \ref{S-results}
describes the effects of the various changes in the analysis.  Section \ref{S-discussion}
discusses these results and gives prospects for the future.
This work elaborates on and updates our 
earlier work on the subject \cite{improvements}.

\section{Data}
     \label{S-Data} 

The line-of-sight velocity data were initially \cite{schou99} analyzed in 74 timeseries of length 72
days, beginning 1 May 1996 00:00:00\_TAI.  The last data point used was
at 12 April 2001 23:20:00\_TAI. In late June 1998, however, contact with SOHO
was lost, resulting in a gap of more than 108 days.  This was followed by
a period of about two months of usable data at the end of 1998, and then
another gap of more than 36 days.  Therefore the 12th timeseries is offset
from the others by 36 days and begins $12 \times 72+36=900$ days after the first,
while the 13th timeseries begins $14 \times 72=1008$ days after the first, as
shown in Table~\ref{T-data} (note the low duty cycles around MDI mission day number 2116). We
have reanalyzed these same 74 time intervals, as well as used them to make 
360-day long timeseries.  Therefore only three of the 72-day long
timeseries were used to make the third 360-day long timeseries, and the last 72-day long
timeseries was unused in the 360-day analysis. Timeseries (and other
final data products) are
available for download from Stanford's Joint Science Operation Center (JSOC). 
See the appendix for details.

To see the effect of the various changes in the processing, we apply them one by one to
the analysis of 15 timeseries covering a period of three years beginning
on 8 January 2004 00:00:00\_TAI.  This is long enough to see an annual component in the
$f$-mode frequencies, but short enough to approximate the solar-cycle
variation as linear during its declining phase. Beginning with the image-scale 
correction, we then apply, in order, corrections for the cubic distortion
from the instrument optics, the misalignment of the CCD, the inclination
error, and the suspected CCD tilt. These are all the corrections that we made during the
spherical harmonic decomposition, and we regenerate timeseries for the
entire mission with all of them applied.  The next two improvements
applied are to the detrending and then the gapfilling.  Again, detrended
and gapfilled timeseries have been regenerated for the entire mission. For
the 360-day analysis, the timeseries were created by
concatenating the detrended and gapfilled 72-day long timeseries.  The
remaining changes to the processing all take place in the fitting.  
We first take into account the horizontal component of the displacement, 
and then distortion of eigenfunctions by the differential rotation
(known as the ``Woodard effect'', see Section \ref{S-leakage}). 
Mode parameters for the entire
mission have been recomputed with these applied, 
using first a symmetric mode profile and again using
an asymmetric one.  This sequence of corrections is 
summarized in Table \ref{T-corrlist}.

\begin{table}

\caption{Timeseries used. Day numbers are given relative to the MDI epoch
of 1 January 1993 00:00:00\_TAI.  Both these and the dates refer to the first
day of the timeseries, and all timeseries begin on the first minute of the day.
Duty cycles are given for the original timeseries
(DC0), the final timeseries (DC1), and the final timeseries after gapfilling
(DC2).  The difference DC0$-$DC1 tends to be positive at the beginning of
the mission (at most $0.031$) and negative at the end (not less than $-0.02$).}

\label{T-data}
\begin{tabular}{ccccc|ccccc}     
%\begin{tabular}{rrrrr|rrrrr}     
\hline
%\multicolumn{2}{c}{<>}
%Day    & Date &  & Duty Cycles &   & Day    & Date &  & Duty Cycles & \\
%Number &      & old & new & filled & Number &      & old & new & filled \\
Day & Date & DC0 & DC1 & DC2 & Day & Date & DC0 & DC1 & DC2 \\
\hline
1216 & 01 May 1996 & 0.895 & 0.888 & 0.907 & 4024 & 08 Jan 2004 & 0.986 & 0.991 & 1.000 \\
1288 & 12 Jul 1996 & 0.964 & 0.949 & 0.966 & 4096 & 20 Mar 2004 & 0.782 & 0.770 & 0.858 \\
1360 & 22 Sep 1996 & 0.964 & 0.954 & 0.969 & 4168 & 31 May 2004 & 0.897 & 0.898 & 0.989 \\
1432 & 03 Dec 1996 & 0.976 & 0.962 & 0.982 & 4240 & 11 Aug 2004 & 0.853 & 0.852 & 0.941 \\
1504 & 13 Feb 1997 & 0.952 & 0.950 & 0.964 & 4312 & 22 Oct 2004 & 0.969 & 0.968 & 0.981 \\
1576 & 26 Apr 1997 & 0.981 & 0.981 & 1.000 & 4384 & 02 Jan 2005 & 0.991 & 0.991 & 1.000 \\
1648 & 07 Jul 1997 & 0.970 & 0.976 & 0.986 & 4456 & 15 Mar 2005 & 0.991 & 0.992 & 0.996 \\
1720 & 17 Sep 1997 & 0.973 & 0.965 & 0.976 & 4528 & 26 May 2005 & 0.983 & 0.989 & 1.000 \\
1792 & 28 Nov 1997 & 0.979 & 0.982 & 1.000 & 4600 & 06 Aug 2005 & 0.989 & 0.988 & 0.996 \\
1864 & 08 Feb 1998 & 0.969 & 0.968 & 0.976 & 4672 & 17 Oct 2005 & 0.985 & 0.985 & 0.996 \\
1936 & 08 Apr 1998 & 0.884 & 0.883 & 0.896 & 4744 & 28 Dec 2005 & 0.988 & 0.992 & 1.000 \\
2116 & 18 Oct 1998 & 0.731 & 0.726 & 0.737 & 4816 & 10 Mar 2006 & 0.990 & 0.992 & 1.000 \\
2224 & 03 Feb 1999 & 0.894 & 0.885 & 0.894 & 4888 & 21 May 2006 & 0.962 & 0.971 & 0.978 \\
2296 & 16 Apr 1999 & 0.982 & 0.974 & 0.986 & 4960 & 01 Aug 2006 & 0.988 & 0.992 & 1.000 \\
2368 & 27 Jun 1999 & 0.986 & 0.987 & 1.000 & 5032 & 12 Oct 2006 & 0.990 & 0.991 & 1.000 \\
2440 & 07 Sep 1999 & 0.930 & 0.917 & 0.941 & 5104 & 23 Dec 2006 & 0.895 & 0.900 & 0.907 \\
2512 & 18 Nov 1999 & 0.870 & 0.839 & 0.852 & 5176 & 05 Mar 2007 & 0.976 & 0.977 & 0.986 \\
2584 & 29 Jan 2000 & 0.986 & 0.983 & 0.989 & 5248 & 16 May 2007 & 0.985 & 0.984 & 0.994 \\
2656 & 10 Apr 2000 & 0.994 & 0.994 & 1.000 & 5320 & 27 Jul 2007 & 0.988 & 0.991 & 1.000 \\
2728 & 21 Jun 2000 & 0.988 & 0.988 & 0.996 & 5392 & 07 Oct 2007 & 0.965 & 0.968 & 0.980 \\
2800 & 01 Sep 2000 & 0.986 & 0.984 & 0.995 & 5464 & 18 Dec 2007 & 0.985 & 0.987 & 1.000 \\
2872 & 12 Nov 2000 & 0.947 & 0.937 & 0.945 & 5536 & 28 Feb 2008 & 0.996 & 0.996 & 1.000 \\
2944 & 23 Jan 2001 & 0.985 & 0.986 & 1.000 & 5608 & 10 May 2008 & 0.989 & 0.993 & 1.000 \\
3016 & 05 Apr 2001 & 0.990 & 0.990 & 1.000 & 5680 & 21 Jul 2008 & 0.988 & 0.991 & 1.000 \\
3088 & 16 Jun 2001 & 0.964 & 0.961 & 0.975 & 5752 & 01 Oct 2008 & 0.983 & 0.986 & 0.994 \\
3160 & 27 Aug 2001 & 0.991 & 0.991 & 1.000 & 5824 & 12 Dec 2008 & 0.983 & 0.989 & 1.000 \\
3232 & 07 Nov 2001 & 0.971 & 0.970 & 0.979 & 5896 & 22 Feb 2009 & 0.996 & 0.996 & 1.000 \\
3304 & 18 Jan 2002 & 0.859 & 0.862 & 0.870 & 5968 & 05 May 2009 & 0.951 & 0.954 & 0.960 \\
3376 & 31 Mar 2002 & 0.987 & 0.985 & 1.000 & 6040 & 16 Jul 2009 & 0.709 & 0.729 & 0.736 \\
3448 & 11 Jun 2002 & 0.978 & 0.984 & 0.996 & 6112 & 26 Sep 2009 & 0.985 & 0.989 & 0.996 \\
3520 & 22 Aug 2002 & 0.991 & 0.990 & 1.000 & 6184 & 07 Dec 2009 & 0.989 & 0.993 & 1.000 \\
3592 & 02 Nov 2002 & 0.994 & 0.994 & 1.000 & 6256 & 17 Feb 2010 & 0.992 & 0.993 & 1.000 \\
3664 & 13 Jan 2003 & 0.992 & 0.989 & 1.000 & 6328 & 30 Apr 2010 & 0.988 & 0.995 & 1.000 \\
3736 & 26 Mar 2003 & 0.982 & 0.982 & 0.996 & 6400 & 11 Jul 2010 & 0.952 & 0.961 & 0.971 \\
3808 & 06 Jun 2003 & 0.822 & 0.826 & 0.852 & 6472 & 21 Sep 2010 & 0.879 & 0.881 & 0.929 \\
3880 & 17 Aug 2003 & 0.981 & 0.981 & 0.996 & 6544 & 02 Dec 2010 & 0.732 & 0.744 & 0.753 \\
3952 & 28 Oct 2003 & 0.878 & 0.878 & 0.952 & 6616 & 02 Feb 2011 & 0.812 & 0.812 & 0.822 \\

\hline
\end{tabular}
\end{table}

\section{Method}
     \label{S-method}

Analysis proceeds as follows.  An observed oscillation mode is taken as 
proportional to the real part of a spherical harmonic given by 
$Y_\ell^m(\phi,\theta) = P_\ell^m(\cos\theta){\rm e}^{{\rm i}m\phi}$, where the $P_\ell^m$ are 
associated Legendre functions normalized such that
\begin{equation} \label{E-plm}
\int_{-1}^1 [P_\ell^m(x)]^2 {\rm d}x =1
\end{equation}
and with the property that $P_\ell^{-m}=P_\ell^m=P_\ell^{|m|}$.  As used here, $\ell$ and $m$ are integers
with $\ell \geq 0$ and $-\ell \leq m \leq \ell$.  However, since 
spherical harmonics with negative $m$ are 
the complex conjugates of those with positive $m$, we
only compute coefficients for $m \geq 0$.  For
medium-$\ell$ analysis, we use degrees up to $\ell=300$.  Beyond this, peaks
along the $f$-mode ridge begin to blend into each other.  For the
$p$-modes, this is already happening around $\ell=200$ or below.

To efficiently compute the spherical harmonic coefficients, each image is remap\-ped to a
uniform grid in longitude and sin(latitude) using a cubic convolution
interpolation, and apodized with a cosine in fractional image radius
from 0.83 to 0.87. The grid rotates at a constant rate of 1/year so that 
the apparent rotation rate of the Sun remains constant. 
The resulting map is Fourier transformed in longitude
and for each $m$ a scalar product is taken with a set of associated Legendre functions of
sin(latitude), which yields the complex amplitudes of the spherical
harmonics as a function of $\ell$ and $m$ in the ranges given above.  These
amplitudes are arranged into timeseries 72 days long, and the timeseries
for each $\ell$ and $m$ is detrended, gapfilled, and Fourier transformed, at
which point the positive frequency part of the transform is identified
with negative $m$ and the conjugate of the negative frequency part is
identified with positive $m$.  The Fourier
transforms are fit (a process that has become known as peakbagging),
resulting in a mode frequency, amplitude, linewidth, and background for
each $\ell$ and $n$.  The $m$-dependence of the frequencies is
parameterized by the $a$-coefficients, which are fit for directly in the
peakbagging, with the other mode parameters assumed to be the same for all
$m$.

Because of leakage between the modes, predominantly caused 
both by projection onto the line of sight and by our
inability to see most of the Sun, the Fourier transform of the target $\ell$
and $m$ contain peaks from neighboring modes as well, which have to be
accounted for in the peakbagging.  This is done through the so-called
leakage matrix, which quantifies the amplitude of each mode as it appears in
the observed spectra. The leakage matrix is calculated by generating artificial images
containing spherical harmonics and their relevant horizontal derivatives, projected onto the line of sight, 
and decomposing them in the same way as the
the actual data.  The same leakage matrix has been used for all times (see
Section \ref{S-pkbgn}).

\subsection{Spherical Harmonic Transform}
     \label{S-sht}

Since spherical harmonic decomposition begins with a remapping, it gives us an
opportunity to apply certain corrections to the data.  The most
significant of these is to correct for the image scale, which is the number 
of arcseconds corresponding to each pixel of the detector.  
Although assumed to be a
constant in the original analysis, changes in the instrument with
temperature and over time actually caused it to vary. 
The radius of the solar image on the MDI detector, measured in pixels, 
is assumed to be given by $\arcsin(D/R_{\rm ref})$ 
divided by the image scale, where $D$ is the observer
distance and $R_{\rm ref}$ is defined as 696 Mm.
Hence the
original value used for the solar radius in pixels was in error. In the
current analysis the image scale is given by a multiplicative factor times
the original constant image scale of 1.97784 arcsec per pixel.  The
inverse of this factor (hence the radius correction) is given as a
function of time $t$ by 
\begin{equation}
f(t) = b_0 + D[b_1 + b_2 (t-t_0) + b_3 (t-t_0)^2].
\end{equation}
The parameters 
$b_0$, $b_1$, $b_2$, $b_3$, and $t_0$ result from a fit to $(A_{\rm major} +
A_{\rm minor})/(2R_0)$, where $A_{\rm major}$ and $A_{\rm minor}$ are the lengths of the major and minor
axes of the solar image returned by the routine used to fit the solar limb
and $R_0$ is the original value used for the solar radius in pixels
(Keh-Cheng Chu, private communication, 2001).  The parameters of the fit change
throughout the mission, typically at a focus change.  Hence the radius
correction is a piecewise-continuous function.

To account for distortion from the instrument optics, we apply a correction
given by an axisymmetric cubic distortion model \cite{korzennik04}.
Such a model gives the distorted coordinate as a cubic function of the undistorted one.  In our 
implementation, the fractional change in coordinates
is given by $C_{\rm dist}(r^2 - R^2)$, where $r$ is the distance from the
center of the CCD, $R$ is the (updated) radius of the solar image, and all
quantities are given in terms of full-disk pixels.  For $C_{\rm dist}$ we have
used $7.06 \times 10^{-9}$, which was derived from a ray-trace of the MDI
instrument. This differs from the value used by
\inlinecite{korzennik04}, which resulted from a different model.  It is
unclear how to resolve the discrepancy, but ongoing
investigation of the MDI distortion is likely to help.

For $P_{\rm eff}$ and $B_0$
we apply a simple sinusoidal correction with respect to time.  Since the
error of the ascending node position is not significant \cite{beckgiles}, if $\delta I$ is the
error of the inclination and $\delta P$ is the error on $P_{\rm eff}$ 
resulting from misalignment of the CCD, then the new values are given by
\begin{equation}
B_0^\prime = B_0 + \delta I \sin[2\pi(t_{\rm obs} - t_{\rm ref})] 
\end{equation}
and
\begin{equation}
P_{\rm eff}^\prime = P_{\rm eff} + \delta P + \delta I \cos[2\pi(t_{\rm obs} - t_{\rm ref})] 
\end{equation}
where $t_{\rm obs}$ is the observation time and $t_{\rm ref}$ is a time when $B_0$ is
zero, both measured in years. For $t_{\rm ref}$ we have used
6 June 2001 06:57:22\_TAI.  For the value of $\delta P$ we have used
$-0.2^\circ$, which agrees with values obtained both by cross-correlations
with GONG images and from the Mercury transit in November 1999 (Cliff Toner,
private communication, 2004).  For the value of $\delta I$ we have used
$-0.1^\circ$, a value derived by \inlinecite{beckgiles}.

The ellipticity of the observed solar image is much greater than the
actual ellipticity of the Sun.  A possible explanation is that
the CCD is not perpendicular to the optical axis of the instrument.  To
correct for this image distortion, we follow the prescription given in the appendix of
\inlinecite{korzennik04}.  The required parameters are the amount [$\beta$]
to rotate the $x$-axis to give the direction around which the CCD is tilted, 
the amount of the tilt [$\alpha$], and the effective focal length [$f_{\rm eff}]$.  
We have adopted the values $\beta = 56.0^{\circ}$, $\alpha =
2.59^{\circ}$, $f_{\rm eff} = 12972.629$ pixels, which are consistent with the
values found by the above-mentioned authors.  Although there is some doubt as
to whether the CCD is actually tilted, the model still reproduces the
observed ellipticity reasonably well (see \opencite{korzennik04}).

\subsection{Detrending and Gapfilling}
     \label{S-dtgf}

Once the 72-day long timeseries have been assembled, the next step in the
processing is the evaluation of the window function.
As used here, the window function is a timeseries of zeros and ones
that identifies both missing data and data that should be rejected on the basis
of quality; only time points corresponding to ones in the window function will
be used in the subsequent processing.  In the original
analysis, the $\ell=0$ timeseries was examined to ensure that gaps resulting
from known spacecraft and instrument events were accurately reflected in
the timeseries generated.  These events included such things as station
keeping, momentum management, problems with the ground antennas, emergency
Sun reacquisitions (ESRs), and tuning changes due to instrumental drifts.  
Additionally, any day whose duty cycle was less than 95\,\% was investigated
to ensure that all potentially available data were processed in the spherical
harmonic decomposition step.  Unfortunately, the original analysis employed a
simple algorithm that performed detrending of the timeseries on full
mission days only, thus requiring any day that contained a discontinuity
in the data, 
such as those caused by instrument tuning changes, to have its window
function zeroed to the nearest day boundary.  Also, the instrument
occasionally stopped taking images, which caused thermal transients after 
it restarted until equilibrium was reestablished.  These turn-on transients,
and other data deemed unusable, were also manually identified in the
timeseries and set to zero in the window function.  Then, ten
timeseries were examined and thresholds on acceptable values in them were
set by hand in order to reject outliers.  These ten timeseries are the real
parts of $\ell=0,m=0$; $\ell=1,m=0$; $\ell=1,m=1$; the imaginary part of $\ell=1,m=1$;  
and the sum over $m$ of the real part squared plus the imaginary part
squared for $\ell$ = 1, 2, 5, 10, 20, and 50.

In the new analysis, we use the old $\ell=0$ timeseries, since they had
already been examined, to confirm the legitimacy of any data missing in
the new $\ell=0$ timeseries.  We then automatically set to zero in the window
function any point where the Image Stabilization System (ISS) was off, as
derived from housekeeping data.  Next we form ten timeseries in the
same fashion as the original analysis, but we replace squaring the real and
imaginary parts in the sum over $m$ with taking the absolute value of the
real and imaginary parts, and then subtract a 41 point
running median. This enables us to remove outliers by taking the rms excluding the top and bottom
1\,\% of the data, and rejecting any points that differ from zero by more
than 6.0 times this rms.

In the new analysis, the discontinuities, which were typically caused by
tuning changes, spacecraft rolls, and any event that powered down the
instrument, all had to be identified by hand.  This information has to be
available for the median filtering, and subsequent detrending can now be
done on entire continuous sections of data irrespective of day boundaries.  
Further, the beginning of every section is automatically checked for the
existence of thermal transients in the $\ell=0$ timeseries by fitting a sum
of two decaying exponentials and a constant. We do not fit the
decay constants as part of this check.  Rather, we fit for them only once
and hold them fixed at values of 15 and 60 minutes.  The use of
two exponentials comes from a model of the instrument. The window
function is zeroed wherever the first two terms of the fit 
differ from zero by more than
the rms of the median-subtracted $\ell=0$ timeseries.  Also, by defining
sections, we were able to manually reject any data lying in between the
sections, if such were deemed necessary.  In the new analysis, defining the
sections of continuous data was the only operation that required human
attention, and had to be done only once.

Detrending in the original analysis was performed on whole mission days
(1440 time points) by fitting a Legendre polynomial of degree given by
$2 + N_{\rm span}/300$ where $N_{\rm span}$ is the number of minutes spanned by the available data 
and the division truncates to the next lowest integer. This
polynomial was subtracted prior to gapfilling, which was also
independently performed on each mission day.  The algorithm used would
compute an autoregressive model from the data and use it to fill gaps up to a maximum size
of five points.  It required six points either before or after each gap to do so,
regardless of the size of the gap.

Detrending in the new analysis is done by fitting a Legendre polynomial of
degree seven to an interval of data spanning 1600 minutes, which is advanced
by 1440 minutes for each fit.  In other words, the detrending
intervals overlap by 160 points.  The polynomials are stitched together in
the overlap region by apodizing each of them with a $\cos^2$ curve. In the 
case that the data points in a detrending interval spanned less than 800 minutes, 
the Legendre polynomial was recomputed for the shorter span, and the fit 
was not apodized. The resulting function is subtracted from the data to 
yield a timeseries with a mean of zero.

In the new analysis, gaps are filled using an autoregressive algorithm
based on the work of \inlinecite{fahlmanulrych}.  This method predicts
values for the missing data based on the spectral content of the data
present. Each point in the known data is expressed as a linear combination
of the $N$ preceding and following points, where $N$ is the order of the
autoregressive model, the coefficients of which are found by minimizing
the prediction error in the least-squares sense.  Hence, the order of the model can be no greater
than the number of points in the shortest section of data.  If a model of
a certain order is desired, it imposes a lower limit on the length of data
sections that can be used to generate it.  In our implementation, we always use
the highest order such that at least 90\,\% of the data will be used to
generate the model, up to a maximum order of 360. 
It was found that increasing the model order beyond this value did not result
in significantly better predictions\footnote{Since the coefficients of a 
model of order $N$ are determined from a model of order $N-1$, our algorithm 
may truncate the model if the ratio of the prediction error to the variance of the timeseries drops below 
$\sim\!1.2 \times 10^{-6}$ as the model order is increased. However, this never occurred
while gapfilling the MDI dataset.}. Once the model is
known, the gaps are filled by minimizing the prediction error in the least-squares 
sense, this time with respect to the unknown data values.
The innovation over the method of Fahlman and Ulrych is that all gaps shorter 
than the model order within each filling interval are filled simultaneously. 
Gaps longer than the model order are not filled. 
Gaps at the beginning or end of the timeseries are not filled regardless of 
their length, because the choice was made not to extrapolate the timeseries.
The model order may possibly then be increased by using the filled values as known 
data, and the process is repeated, but using the original gap structure.  That is, 
the gaps that were filled on the first iteration will be filled again using the 
new model.  If the model order did not change, or if all the gaps were already filled 
in the first iteration, the process stops after two iterations.  Otherwise a final
iteration is run wherein a new model is computed using the newly filled values, and 
the gaps are filled one last time (Rasmus Larsen, private communication, 2013).
Lastly, a new window function is generated to reflect the filled gaps.

\subsection{Peakbagging}
     \label{S-pkbgn}

Fourier transforms of the gapfilled timeseries are fit using a maximum-likelihood 
technique, taking into account leakage between the modes. 
In this section we expand upon the presentation given by \inlinecite{schouthesis} 
and describe the fitting process as it is currently implemented.
When modelling an oscillation mode as a stochastically excited damped
oscillator, both the real and imaginary parts of the Fourier transform
will be normally distributed with a mean of zero. The variance due to the
mode will be given by
\begin{equation} \label{E-var}
v(\nu_0,w,A,\nu) = \frac{2wA^2}{w^2+4(\nu-\nu_0)^2}
\end{equation}
where $\nu_0$ is the frequency of the mode, $w$ is the full width at half
maximum, and $A$ is the amplitude ($A^2$ is a measure of the total power
in the mode). To fit an actual observed spectrum, one must also add a
background term; our treatment of the background is described below. 
Furthermore, to account for the redistribution of power caused by gaps in the
timeseries, this model will be convolved with the power spectrum of the window function
\cite{anderson90}.
If $x$ is the real part of the observed value of the Fourier transform, then
the probability density for the $i$th frequency bin in the real part will
by given by
\begin{equation} \label{E-probr}
P_{\rm real}(\nu_0,w,A,\nu_i) = 
\frac{1}{\sqrt{2\pi v(\nu_i)}} \exp\left(-\frac{x(\nu_i)^2}{2v(\nu_i)}\right)
\end{equation}
and likewise for $P_{\rm imag}$ with $x$ replaced by $y$, the imaginary part.
The total probability density for the $i$th bin is then $P=P_{\rm real}P_{\rm imag}$.
In these equations the mode parameters, and hence $v$, are functions of $n$, $\ell$, and $m$; 
we have suppressed their dependence on these for conciseness.

The idea behind the maximum-likelihood approach is to maximize the joint probability density of a given mode,
which is given by a product of individual probability densities over a
suitable number of frequency bins (assuming that each frequency bin is
independent, which is not strictly true in the presence of gaps). This is
equivalent to minimizing the negative logarithm of this product, which,
except for constants, is given by
\begin{equation} \label{E-min}
S(\nu_0,w,A) = \sum_i \ln \left(v(\nu_i)\right) + 
\frac{x(\nu_i)^2+y(\nu_i)^2}{v(\nu_i)}.
\end{equation}
where $\nu_i$ is the frequency of the $i$th frequency bin.
For a given value of $\ell$, there will be $2\ell+1$ values of $m$. Rather than
fitting each $m$ separately, we will maximize the joint probability
density of all of them together.  To do so, we assume that the width and amplitude 
are independent of $m$ and estimate the variation of the background with $m$ 
from the spectrum far from the peaks.  We redefine $\nu_0$ as the mean multiplet
frequency for each $n$ and $\ell$, and expand the frequency dependence on $m$ as 
\begin{equation} \label{E-acoeff}
\nu_{n\ell m} = \nu_0(n,\ell) + \sum_{i=1}^{N_a} a_i(n,\ell) \mathcal{P}_i^\ell(m)
\end{equation}
where the polynomials [$\mathcal{P}$] are those used by
\inlinecite{schou94inv}, and the coefficients [$a_i$] are fit for directly.
The $a_1$ coefficient will have 31.7\,nHz added to correct for the average orbital 
frequency of the Earth about the Sun.
In what follows, we will label
the set of parameters upon which $S$ depends using the vector
$\mathbfit{p}$.  This will include $\nu_0$, $w$, $A$, $N_a$
$a$-coefficients, a background parameter (described below), and optionally
a parameter to describe the asymmetry (also described below), for each $n$
and $\ell$.

Due to leakage between the modes, the observed timeseries and Fourier
transforms are a superposition of the true underlying oscillations.  The
observed timeseries for a given $\ell$ and $m$ will be given by
\begin{equation} \label{E-tsleaks}
o_{\ell m}(t) = \sum_{n^\prime \ell^\prime m^\prime} c_{\ell m,\ell^\prime m^\prime}^{RR} 
\mathrm{Re}[a_{n^\prime \ell^\prime m^\prime}(t)] + i c_{\ell m,\ell^\prime m^\prime}^{II} 
\mathrm{Im}[a_{n^\prime \ell^\prime m^\prime}(t)]
\end{equation}
where $a(t)$ is the complex amplitude of the underlying timeseries, and
$\mathrm{Re}[\,]$ and $\mathrm{Im}[\,]$ denote the real and imaginary
parts, respectively. The sensitivity coefficients $c^{RR}$ and $c^{II}$
give the real-to-real leaks and imaginary-to-imaginary leaks respectively.  
Approximate expressions for the radial contribution to these coefficients
are given by \inlinecite{schou94leaks}.  Under the same approximations, it
can be shown that the real-to-imaginary and imaginary-to-real leaks are
identically zero for geometries that are symmetric around the central
meridian. Although these are still assumed to be zero for the current
work, $c^{RR}$ and $c^{II}$ are computed as described
below.  It can also be shown under these assumptions that
\begin{eqnarray} \label{E-csym}
c_{\ell m,\ell^\prime m^\prime}^{RR} &=& c_{\ell^\prime m^\prime,\ell m}^{RR} \nonumber \\
c_{\ell m,\ell^\prime m^\prime}^{II} &=& c_{\ell^\prime m^\prime,\ell m}^{II} \nonumber \\
c_{\ell(-m),\ell^\prime m^\prime}^{RR} &=& c_{\ell m,\ell^\prime m^\prime}^{RR} \nonumber \\
c_{\ell(-m),\ell^\prime m^\prime}^{II} &=& -c_{\ell m,\ell^\prime m^\prime}^{II}
\end{eqnarray}
and that $c^{RR}=c^{II}=0$ when $\ell+m+\ell^\prime+m^\prime$ is odd.
Note that since the spherical harmonic decomposition is not able to separate the 
different values of $n$, we have suppressed the $n$-dependence of the leaks in 
these equations. Later we will consider effects that cause the leaks to vary 
with $n$.
In frequency space, the observed Fourier transform can then be expressed as
\begin{equation} \label{E-ftleaks}
\tilde{o}_{\ell m}(\nu) = x_{\ell m}(\nu) + iy_{\ell m}(\nu) = \sum_{n^\prime \ell^\prime m^\prime} 
C_{\ell m,\ell^\prime m^\prime} \tilde{a}_{n^\prime \ell^\prime m^\prime}(\nu)
\end{equation}
where $C = (c^{RR} + c^{II})/2$ (\opencite{schou94leaks}). Although in
principle the sum above should be over all modes, for a given $\ell$ and $m$,
only modes in a certain range in $\ell^\prime$ and $m^\prime$ will have
significant leakage.  Therefore the sum in Equation (\ref{E-ftleaks}) need only be over modes that
may have appreciable amplitudes within the fitting window.  For this work
we have used $\Delta \ell = \ell - \ell^\prime$ in the range $\pm 6$ and $\Delta m
= m - m^\prime$ in the range $\pm 15$. Furthermore, we neglect leaks for 
$\Delta \ell + \Delta m$ odd or which are estimated to be far away in frequency. 
Since the modes on the Sun are
uncorrelated with each other, the elements of the covariance matrix
between the different transforms at each frequency point will be given by
\begin{eqnarray} \label{E-cov}
E_{\ell m,\ell^\prime m^\prime}^{\rm modes}(\nu_i) &=& 
{\rm Cov}[x_{\ell m}(\nu_i),x_{\ell^\prime m^\prime}(\nu_i)] = 
{\rm Cov}[y_{\ell m}(\nu_i),y_{\ell^\prime m^\prime}(\nu_i)] \nonumber \\
 &=& \sum_{n^{\prime\prime} \ell^{\prime\prime}m^{\prime\prime}} 
C_{\ell m,\ell^{\prime\prime}m^{\prime\prime}}
C_{\ell^\prime m^\prime,\ell^{\prime\prime}m^{\prime\prime}}
v_{n^{\prime\prime} \ell^{\prime\prime}m^{\prime\prime}}(\mathbfit{p},\nu_i).
\end{eqnarray}
The total covariance will be the sum of the covariance between the modes
and the covariance of the noise. Since we fit each $\ell$ separately and all
$m$ for that $\ell$ simultaneously, the elements of the covariance matrix [$\mathbf{E}$] used in the fitting
are given by
\begin{equation} \label{E-cov2}
E_{m,m^\prime}(\nu_i) = E_{m,m^\prime}^{\rm modes}(\nu_i) + \tilde{E}_{m,m^\prime}\frac{\nu_B}{\nu_i}e^b
\end{equation}
where $\tilde{E}_{m,m^\prime}$ is the measured covariance between $m$ and $m^\prime$ in
the frequency range 7638.9 to 8217.6\,$\mu$Hz, $\nu_B$ is a constant, and $b$ is a free
parameter determined in the fit. Due to our choice of normalization, $e^b$
is proportional to the length of the timeseries. The probability density
for a frequency bin then becomes
\begin{equation} \label{E-prob2}
P(\mathbfit{p},\nu_i) =  \frac{1}{|2\pi\mathbf{E}(\mathbfit{p},\nu_i)|}
\exp\left[-\frac{1}{2}(\mathbfit{x}(\nu_i)^T\mathbf{E}(\mathbfit{p},\nu_i)\mathbfit{x}(\nu_i) + \mathbfit{y}(\nu_i)^T\mathbf{E}(\mathbfit{p},\nu_i)\mathbfit{y}(\nu_i)) \right] 
\end{equation}
and the function to minimize becomes 
\begin{equation} \label{E-min2}
S(\mathbfit{p}) = \sum_i \ln|\mathbf{E}(\mathbfit{p},\nu_i)| + 
\mathbfit{x}(\nu_i)^T\mathbf{E}(\mathbfit{p},\nu_i)\mathbfit{x}(\nu_i) + \mathbfit{y}(\nu_i)^T\mathbf{E}(\mathbfit{p},\nu_i)\mathbfit{y}(\nu_i)
\end{equation}
where $|~|$ denotes the determinant, $\mathbfit{x}$ is a vector of the
$2\ell+1$ real parts of the transforms, and $\mathbfit{y}$ is a vector of the
$2\ell+1$ imaginary parts.  Note that $\mathbfit{p}$, $\mathbfit{x}$, and
$\mathbfit{y}$ are implicit functions of $n$ and $\ell$ (the dependence of
$\mathbfit{x}$ and $\mathbfit{y}$ on $n$ come from the frequency range
chosen for the fitting window).  For the width of the fitting window we
have chosen 5.0 times the estimated width of the peak, with a minimum of 
2.9\,$\mu$Hz and a maximum of 81.0\,$\mu$Hz. 
The minimum ensures that we always have enough 
points in frequency for the fit to be stable, and the maximum serves to 
limit the computational burden.
The peakbagging will yield the mode parameters specified by
$\mathbfit{p}$ for each multiplet that it is able to fit, as well as error
estimates on these, generically referred to as $\sigma$.  The errors are
estimated from the inverse of the Hessian matrix at the minimum of $S$.
For readability, the error estimates for the $a$-coefficients will
be labelled by $\sigma_i$, while the rest will be designated in the usual way.

The minimization scheme used is a variation of the Levenberg--Marquardt
method.  For further details, such as approximations made in the calculation
of derivatives, the reader is referred to \inlinecite{schouthesis}.

Since we fit for one $n$ and $\ell$ at a time while holding the leaks fixed,
the peakbagging must be iterated to account for the variation of the mode parameters 
of the leaks
as the fits proceed. For all iterations except the last, we fit six $a$-coefficients. 
In the original analysis, the initial guess for the
first iteration was taken from the final fits of the previous timeseries.  
In the new analysis, the same initial guess was used for all time periods,
which allows for fitting all of them independently of one another.  We
found this made no significant difference.  Any modes that cannot be fit
in the first attempt have the initial guess of their background parameter [$b$]
perturbed by $-1$ and the fit is reattempted.  At this point in the original
analysis the resulting set of fitted modes would be weeded by hand to
reject outliers.  In the new analysis this step is simply skipped; again
we found it made no significant difference.  In both cases the remaining
modes are used to make new initial guesses for the modes that had not
converged (or were rejected).  The second iteration is then done in the
same way as the first. At no point do we ever attempt to fit modes for
which there are estimated to be other modes within $\pm 2$ in $\ell$ and
within twice the line width in frequency.  These typically occur at the ends of ridges
and do not converge in any case.

For subsequent iterations, the modes that have not converged to within
$0.1\,\sigma_{\nu_0}$ or for which there exist unconverged modes with the same
$n$ and $\Delta \ell=\pm 1$ are fitted (occasionally more modes would
be fit in the original analysis). In the original
analysis the convergence of the modes would be examined to determine the
total number of iterations, which would usually be from 9 to 11.  All
modes would be fit in the last iteration and in at least 
one of the preceding two iterations.  In the new analysis,
for the sake of automation, the peakbagging would always be performed for
ten iterations with all modes being fit during the last three.  In both
cases, the final fits are repeated with both 18 and 36 $a$-coefficients, 
which is to say that these fits are not iterated.

After the final iteration, the resulting set of modes is automatically
weeded one last time. For the fits with six $a$-coefficients, modes
differing by more than $0.25\,\sigma_{\nu_0}$ from their input guesses 
are rejected.  Additionally, any mode with a large error on its
frequency given its width is suspect: if there were no background noise,
we would expect a frequency error given by
\begin{equation} \label{E-idealsig}
(2\ell+1)\sigma^2_{\nu_0} = \frac{w}{4\pi T}
\end{equation}
where $T$ is the length of the timeseries (\opencite{libbrecht92}).  Any
mode with a frequency error greater than 6.0 times this prediction is
rejected. The same theoretical error estimate is the motivation for
identifying modes for which the line width is smaller than the width of a
frequency bin.  These modes have the error estimates on their frequencies
and $a$-coefficients increased by a factor of $\sqrt{1/(wT)}$. This
prevents underestimates of the error caused by low estimates of the widths
in the region where they cannot be reliably estimated. 

The resulting set
of mode parameters is then compared to those of a model obtained from a rotational
inversion of fits to a 360-day long timeseries at the beginning of the mission.  
The median difference between the fit and the model of the odd
$a$-coefficients is taken for the $f$-modes to account for their change
throughout the solar cycle.  The differences for all of the modes are
compared to this median; any that differ by more than $10\,\sigma$ are
rejected.

To weed the fits with 18 and 36 $a$-coefficients, their error estimates are
adjusted as above.  Frequencies and $a$-coefficients are then compared to
the fits using six $a$-coefficients. Any mode for which the error estimates on any 
of these parameters increased by more than a factor of 2.0,
or for which any of these parameters changed by more than $2\,\sigma$ 
(estimated from the fits with 18 and 36 $a$-coefficients, respectively), is rejected.
Any mode that was rejected in the fits with six $a$-coefficients is
also removed from the fits with 18 $a$-coefficients, and any mode that was
rejected in the fits with 18 $a$-coefficients is also removed from the fits with 36
$a$-coefficients.

\subsubsection{Leakage Matrix}
  \label{S-leakage}

For this work, the leakage matrix elements, which quantify how modes nearby 
in spherical harmonic space appear in the spectrum of the target mode, are computed by generating
artificial images containing components of the solution to the oscillation equations
projected onto the line of sight for a subset of the modes that we wish to fit.  
A mode on the Sun has a velocity at the surface with components proportional to  
the real parts of \footnote{The sign of $u_r$ relative to $u_\theta$ and
$u_\phi$ depends on the convention for the sign of $m$.}
\begin{eqnarray} \label{E-velocity}
u_r &=& Y_\ell^m(\phi,\theta) = P_\ell^m(x){\rm e}^{{\rm i}m\phi} \nonumber \\
u_\theta &=& \frac{1}{L} \pder{Y_\ell^m}{\theta} = 
-\frac{1}{L} \deriv{P_\ell^m}{x}{\rm e}^{{\rm i}m\phi}\sin\theta \nonumber \\
u_\phi &=& \frac{1}{L} \frac{1}{\sin\theta}\pder{Y_\ell^m}{\phi} = 
\frac{1}{L} \frac{{\rm i}m}{\sin\theta}P_\ell^m(x){\rm e}^{{\rm i}m\phi}
\end{eqnarray}
where $x=\cos\theta$ and $L=\sqrt{\ell(\ell+1)}$. A mode with oscillation
amplitude $V_{\ell m}$ will then have a total velocity of 
\begin{equation} \label{E-vtot}
\mathbfit{V} = \mathbfit{V}_{\ell m}^r + c_t\mathbfit{V}_{\ell m}^h
\end{equation}
where $\mathbfit{V}_{\ell m}^r = V_{\ell m}u_r\hat{r}$,
$\mathbfit{V}_{\ell m}^h = V_{\ell m}(u_\theta\hat{\theta}+u_\phi\hat{\phi})$, and 
\begin{equation} \label{E-ct}
c_t = \frac{\nu_0^2(0,\ell)}{\nu_0^2(n,\ell)}
\end{equation}
is the ratio of the mean multiplet frequency of the $f$-mode squared to the 
mean multiplet frequency of
the given mode squared at that $\ell$ \cite{rhodesreiter01}.  Therefore $c_t=1$ for the $f$-mode and
$c_t<1$ for the $p$-modes.
Equation (\ref{E-ct}) is derived under the assumption of zero lagrangian pressure 
perturbation at the solar surface.

Since the spherical harmonic decomposition does not separate the different radial orders, 
we create a separate matrix for the
vertical and horizontal components; the effective leakage matrix will be
computed during the fitting by combining them according to Equation (\ref{E-vtot}). 
We project each component onto the line of sight separately
using projection factors calculated for a finite observer distance.  In
the approximation of an infinite observer distance this would become
\begin{eqnarray} \label{E-projected}
u_{\rm vertical} &=& V_{\ell m} P_\ell^m(x){\rm e}^{{\rm i}m\phi}\sin\theta\cos\phi \nonumber \\
u_{\rm horizontal} &=& -\frac{V_{\ell m}}{L} \left(\deriv{P_\ell^m}{x}\sin\theta\cos\theta\cos\phi
 + \frac{im}{\sin\theta}P_\ell^m(x)\sin\phi\right){\rm e}^{{\rm i}m\phi}
\end{eqnarray}
where we choose $V_{lm} = 1000\ {\rm ms}^{-1}$ to give us roughly the
same order of magnitude as the observations. As with the real data, these
images are only calculated for $m \geq 0$. The resulting leakage matrix 
will be divided by 1000.

These images are first generated as they would appear to MDI, assuming an
observer distance of 1 AU, a $P_{\rm eff}$ and $B_0$ both equal to zero, and that
the image is centered on the CCD.  They are then convolved with a gaussian
in each dimension with the same width of $\sigma = 4/\sqrt{2}$ as used
onboard the spacecraft, but they are not sub-sampled at this point.  
Rather they are also
convolved with a function that takes into account the interpolation errors
made during the subsequent remapping. This function is generated by applying the cubic 
convolution algorithm to a $\delta$-function. During the spherical harmonic decomposition, 
these images will be remapped to the same resolution in longitude and
sin(latitude) as the real data.  The higher resolution images are used to
simulate an average over different pixel offsets; we have verified the
accuracy of this technique by generating lower resolution images and
actually performing the average. After the remap, the artificial data are
processed exactly as the real data.  For each image, we take its scalar
product with a set of target spherical harmonics in the range $\Delta \ell =
\pm 6$ given above.  The results are the coefficients $c^{RR}$ and
$c^{II}$ given in Equation (\ref{E-tsleaks}).  The values for the modes that we
did not compute directly are found by interpolation.  The values for
negative $m$ are given by Equations (\ref{E-csym}).

In the original analysis, only the vertical component of the leakage
matrix was used, meaning that the horizontal component was assumed to be
zero.  Although this is not a bad approximation for high-order $p$-modes, it
becomes worse as one approaches the $f$-mode ridge, where the horizontal
and vertical components have equal magnitude.  In the new analysis, our
first improvement to the peakbagging is to include both components.  

For a spherically symmetric Sun, the horizontal eigenfunctions would be
spherical harmonics.  Although the presence of differential rotation
breaks this symmetry, the true eigenfunctions can still be expressed as a
sum over spherical harmonics.  In the new analysis, this is accounted for
in the peakbagging by appropriately summing the leakage matrix.  We use
the prescription given by \inlinecite{woodard89} with the differential
rotation expanded as
\begin{equation} \label{E-omega}
\Omega(x) = B_0 + B_1 x^2 + B_2 x^4
\end{equation}
where, again, $x=\cos\theta=\sin(\mathrm{latitude})$.  We first used
constants derived from surface measurements, with values of $B_1=-75\,$nHz
and $B_2=-50\,$nHz as given by Woodard (the value of $B_0$ is not used).
However, this has the drawback of distorting every mode in the same way,
even though they sample different depths where the differential rotation
has a different dependence on latitude. Following
\inlinecite{vorontsov07}, we use the estimated splitting coefficients to
calculate $B_1$ and $B_2$ for each mode separately.  In particular, 
we use the approximation that 
\begin{eqnarray} \label{E-woodb}
B_1 &=& -5a_3 - 14a_5 \nonumber \\
B_2 &=& 21a_5
\end{eqnarray}
so that $B_1$ and $B_2$ change as the iteration proceeds. Fortunately this
did not disrupt the convergence of the $a$-coefficients.  This change made
only a modest difference in the mode parameters, as discussed below.

\subsubsection{Asymmetry}
  \label{S-asymmetry}

In addition to the symmetric line profiles described by Equation (\ref{E-var}), 
we have also used asymmetric profiles to fit the data.  Although it is common 
to use the profile derived by \inlinecite{nigam98}, their equation has the 
undesirable properties that it is based on an approximation that does not 
hold far from the mode frequencies and that its integral over all frequencies is infinite. 
  To derive a more 
well behaved profile, we begin with Equation (3) of \inlinecite{nigam98}, which was
derived for a one-dimensional rectangular potential well model, and generalize it by replacing their $\beta X$ 
with an arbitrary function of frequency $h(\nu)$. Since $\beta$ is generally very small,
we drop the second term in the numerator to arrive at a variance given by
\begin{equation}
v(\nu) = \frac{P_D(\nu) \cos^2[h(\nu)+\gamma(\nu)]}{g(\nu)+\sin^2[h(\nu)]}
\label{E-asym1}
\end{equation}
where $P_D$ is the power spectrum of the excitation, $\gamma$ is a measure of the
asymmetry, and $g$ is related to the damping.  The function $h$ is constrained to 
be $n\pi$ at the mode frequencies, and in the numerator we have changed sin to cos so that $\gamma=0$ 
corresponds to a symmetric profile. Considering a single $\ell$ and $m$, we can expand Equation (\ref{E-asym1}) 
in terms of profiles given by Equation (\ref{E-var}) to get 
\begin{equation}
v_{\ell m}(\nu) =
\cos^2[h_{\ell m}(\nu)+\gamma_{\ell m}(\nu)]
\sum_n \frac{1}{\cos^2[\gamma_{\ell m}(\nu_{n\ell m})]}\frac{2wA^2}{w^2+4(\nu-\nu_{n\ell m})^2}
\label{asym2}
\end{equation}
where the factor $1/\cos^2(\gamma_{\ell m}(\nu_{n\ell m}))$ has been included so that to lowest order, 
$A$ retains its original meaning. To find a function to use for $h$, we note that
from the Duvall law \cite{duvall82} we can define 
$h_0(\nu) = \nu F(\nu/(\ell+1/2)) - \pi\alpha(\nu) \approx n\pi $, where $F$ and $\alpha$ 
are known functions.  These we have tabulated from a fit to a 360-day long timeseries at the beginning 
of the mission, and interpolate them as needed during the peakbagging.
We then choose
$h = h_0 + h_1$ where $h_1$ is a piecewise linear function chosen to make 
$h$ exactly $n\pi$ at the mode frequencies as required. The function $\gamma$ can likewise
be interpolated using a piecewise linear function derived from its value at the mode frequencies. 
Above the frequency of the maximum $n$ and below the frequency of the minimum $n$, we 
assign constant values to $h_1$ and $\gamma$.

Equation (\ref{asym2}) is valid for all frequencies. Restricting ourselves to a single mode, 
we can now replace the variance in Equation (\ref{E-cov}) with
\begin{equation}
v_{n\ell m}(\mathbfit{p},\nu) = \frac{\cos^2[h_{\ell}(\nu^\prime)+\gamma_{\ell}(\nu^\prime)]}
{\cos^2(\gamma_{n\ell})}\frac{2w_{nl}A_{n\ell}^2}{w_{n\ell}^2+4(\nu-\nu_{n\ell m})^2}
\label{asym3}
\end{equation}
where $\nu^\prime = \nu - \nu_{n\ell m} + \nu_0(n,\ell)$, $\nu_{n\ell m}$ is given by 
Equation (\ref{E-acoeff}), and we have implicitly assumed that the asymmetry is 
the same for all $m$.  The function $\gamma_\ell$ is constructed from the values 
$\gamma_{n\ell}$, the fit parameters, such that $\gamma_\ell(\nu_0(n,\ell))=\gamma_{n\ell}$.
Since $h$ is an increasing function of frequency, a positive value of $\gamma_{nl}$ means
that the high-frequency wing of the line will be lower than the low-frequency wing.
Finally, the value actually reported is $\tan(\gamma_{n\ell})$.

To form the initial guess for the asymmetric
fits, we examined the frequencies and asymmetry parameters resulting from
a preliminary fit using the same initial guess as for the symmetric fits.  
We then fit the frequency shift relative to the symmetric case by fitting a 
sixth-order polynomial in frequency, which we now add to the initial guess for
the frequency. For the asymmetry parameter, we use a third-degree
polynomial in frequency directly for the initial guess.

When we tried the iteration scheme described above for the 15 intervals that we
analyzed in detail, we found that for some of them very few $f$-modes were
fitted.  We therefore added an automatic rejection of fits with negative
asymmetry parameters in the range $\nu < 2000\,\mu{\rm Hz}$ between iterations
of the peakbagging, since the asymmetry in that range is observed to be positive.  
This solved the problem for these 15 intervals, but
when we reanalyzed the entire mission, a small number of intervals still had
few $f$-modes fit.  We were able to improve the coverage of those intervals by
adding a further criterion to reject modes that had an extremely high value of 
$\tan(\gamma)$, but this caused other intervals to lose modes.  We
therefore reverted to the initial rejection criteria.  Clearly, the
asymmetric fits are much less stable than those using symmetric profiles.

\section{Results}
     \label{S-results} 

\subsection{Mode Parameters}
   \label{S-modeparms}

We applied 11 different analyses to 15 intervals of 72 days each, beginning in January 2004 
(see Table \ref{T-corrlist}). 
Comparing the analyses is complicated by the 
fact that, in general, they do not result in identical modesets. For each analysis,
we therefore only consider modes common with the preceding analysis for each interval.
  We then took an average in
time over whatever intervals had each mode successfully fit.  In so doing,
we are assuming that the difference in mode parameters resulting from the
difference in the analysis is much more significant than their relative
change over time.  In the following figures, we plot the difference in
several mode parameters normalized by their error estimates.  For these
plots, we calculated the average error estimates, rather than the error on
the average, and for any given comparison between two analyses, we use the
larger error estimate of the two.  Thus the significance that we have plotted
is the least that one might expect from a single 72-day fit.  The range of some
plots excludes a few outliers; this is always less than 1.4\,\% of the 
data.  The sense of subtraction is the later analysis minus the earlier one.
Here we have plotted all of the parameter differences as a function of
frequency.  Full listings of all mode parameters for all time intervals and all
analyses that we performed are provided as ASCII tables in the electronic
supplementary material.

\begin{figure}

\centering
{$\Delta\nu_0/\sigma_{\nu_0}$}\par\medskip

\centerline{\includegraphics[width=1.0\textwidth,clip=]{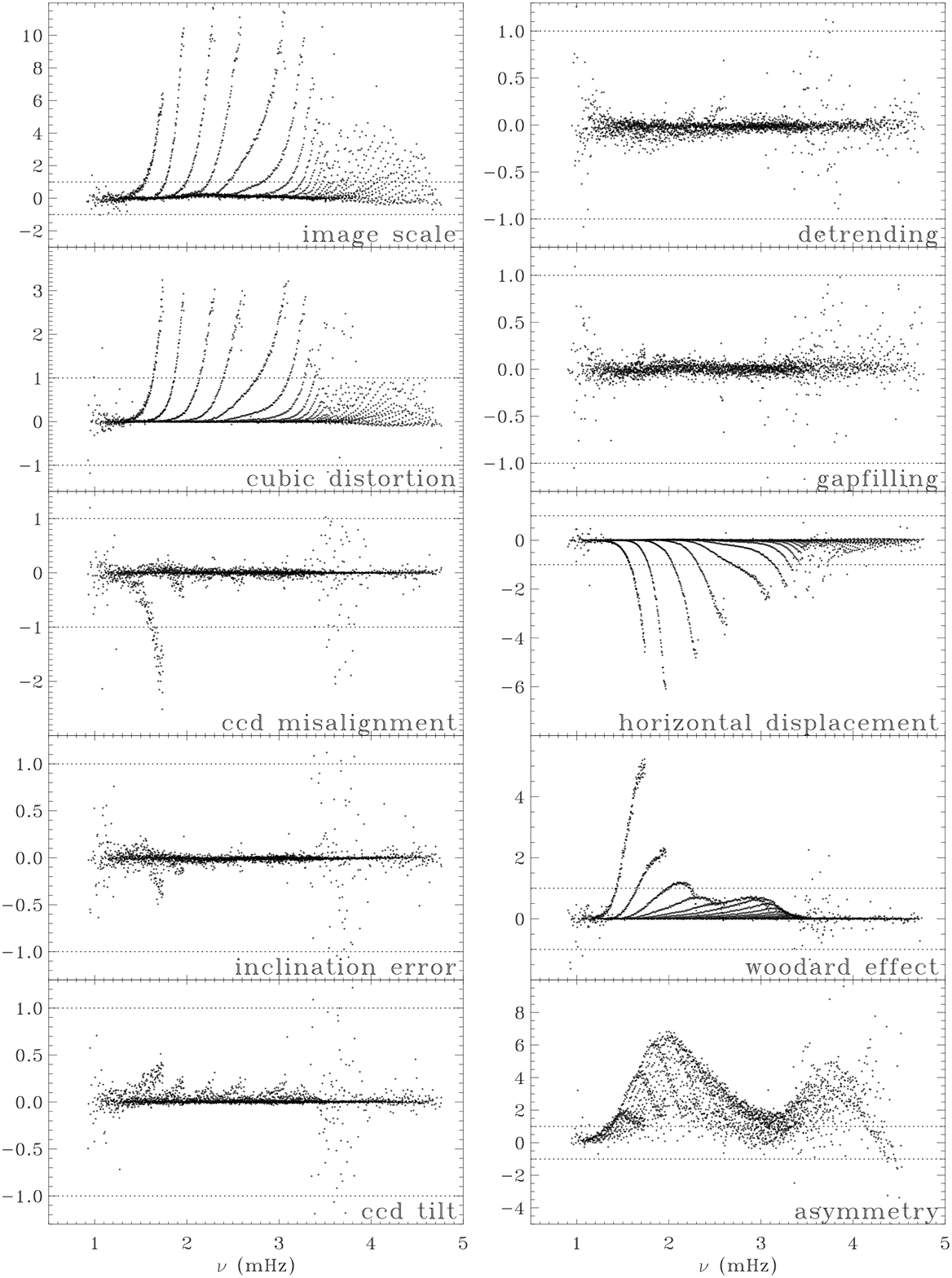}}

\caption{Change in mean multiplet frequency resulting from each change as
a function of frequency, in units of standard deviation. Each panel is scaled differently;
dotted lines show the $\pm 1 \sigma$ levels.}

\label{F-nudiff}
\end{figure}

\begin{figure}

\centering
{$\Delta A/\sigma_A$}\par\medskip

\centerline{\includegraphics[width=1.0\textwidth,clip=]{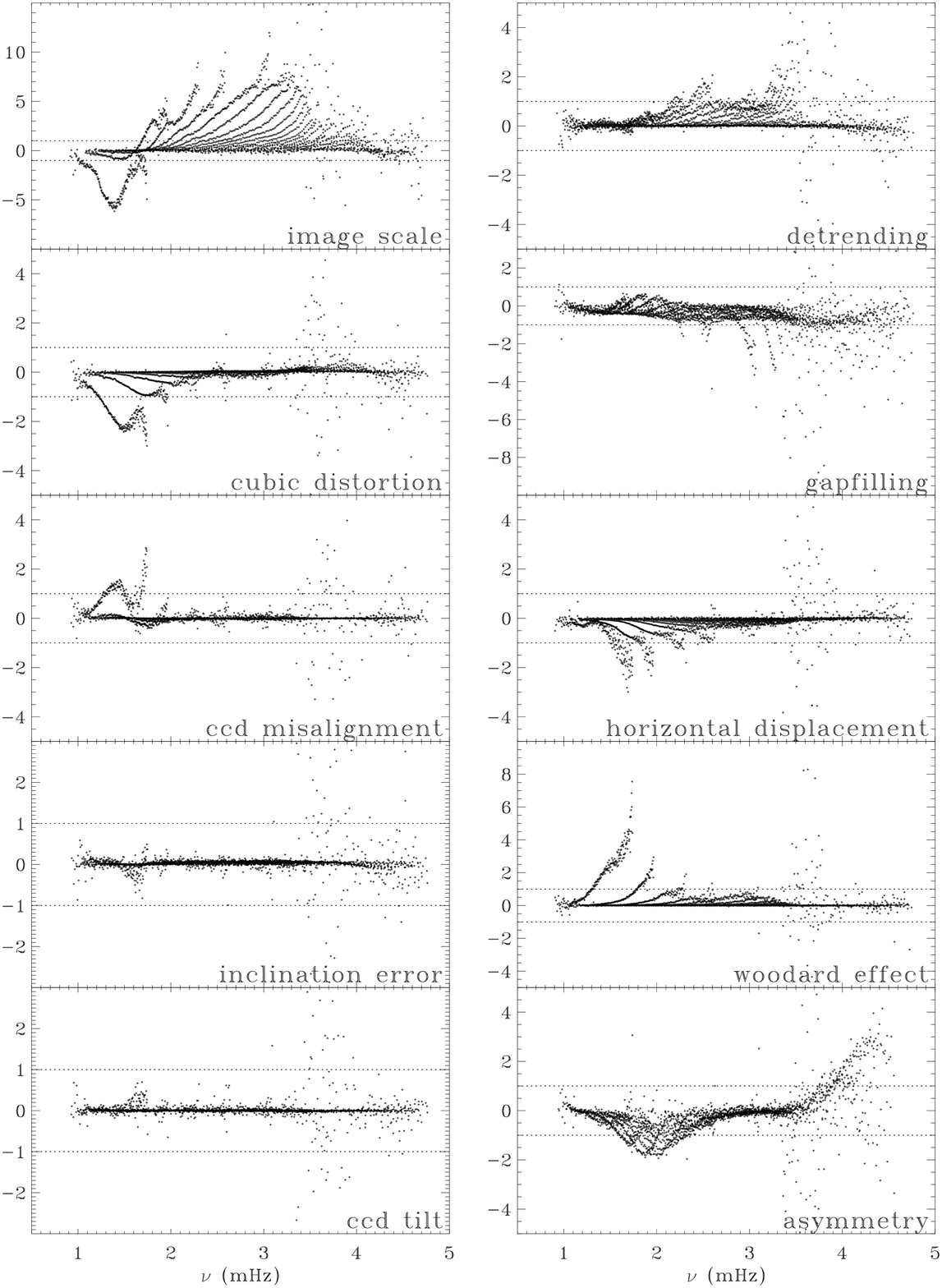}}

\caption{Change in amplitude resulting from each change as a function of
frequency, in units of standard deviation. Each panel is scaled differently;
dotted lines show the $\pm 1 \sigma$ levels.}

\label{F-ampdiff}
\end{figure}

\begin{figure}

\centering
{$\Delta w/\sigma_w$}\par\medskip

\centerline{\includegraphics[width=1.0\textwidth,clip=]{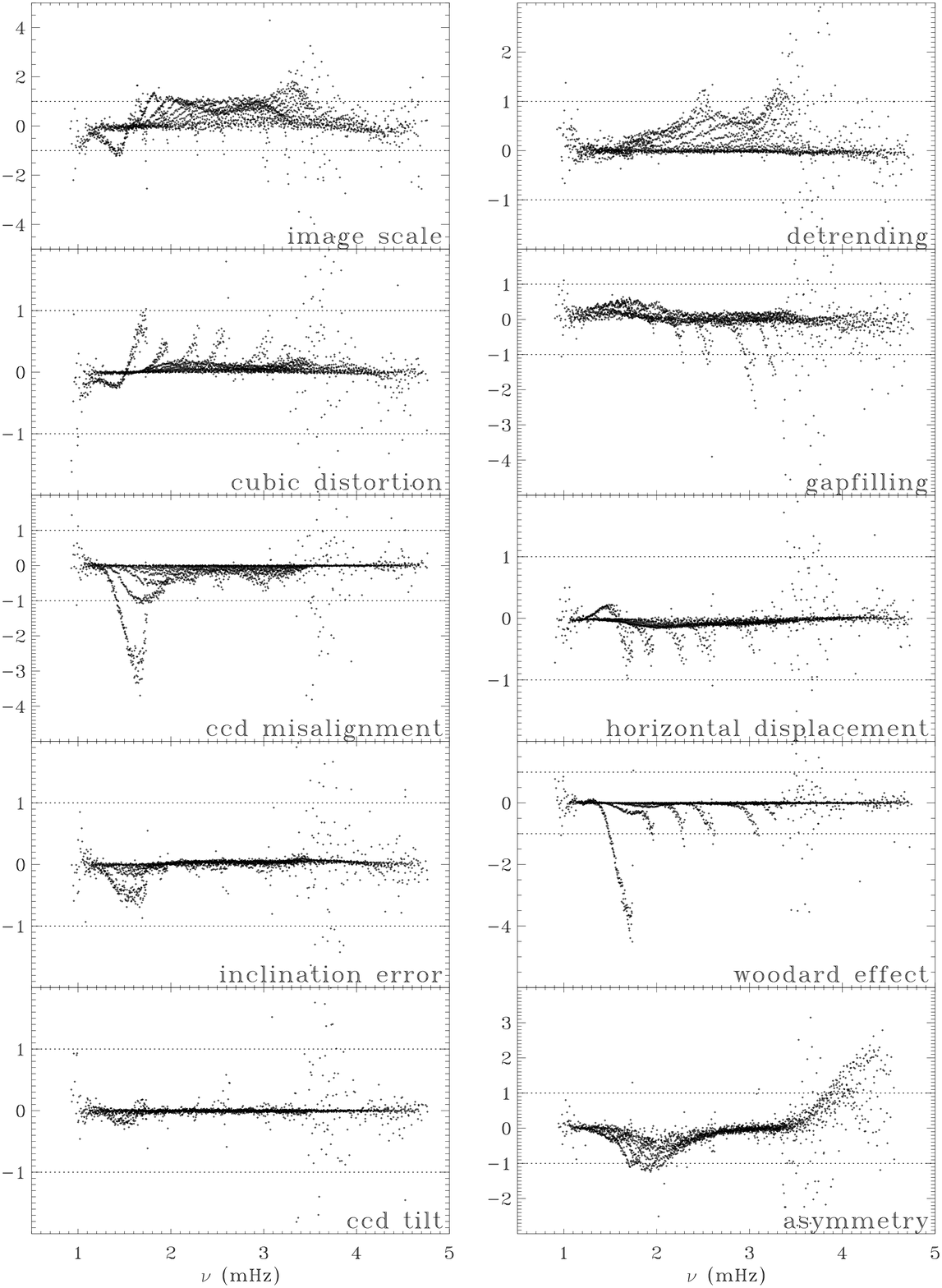}}

\caption{Change in width resulting from each change as a function of
frequency, in units of standard deviation. Each panel is scaled differently;
dotted lines show the $\pm 1 \sigma$ levels.}

\label{F-widdiff}
\end{figure}

\begin{figure}

\centering
{$\Delta b/\sigma_b$}\par\medskip

\centerline{\includegraphics[width=1.0\textwidth,clip=]{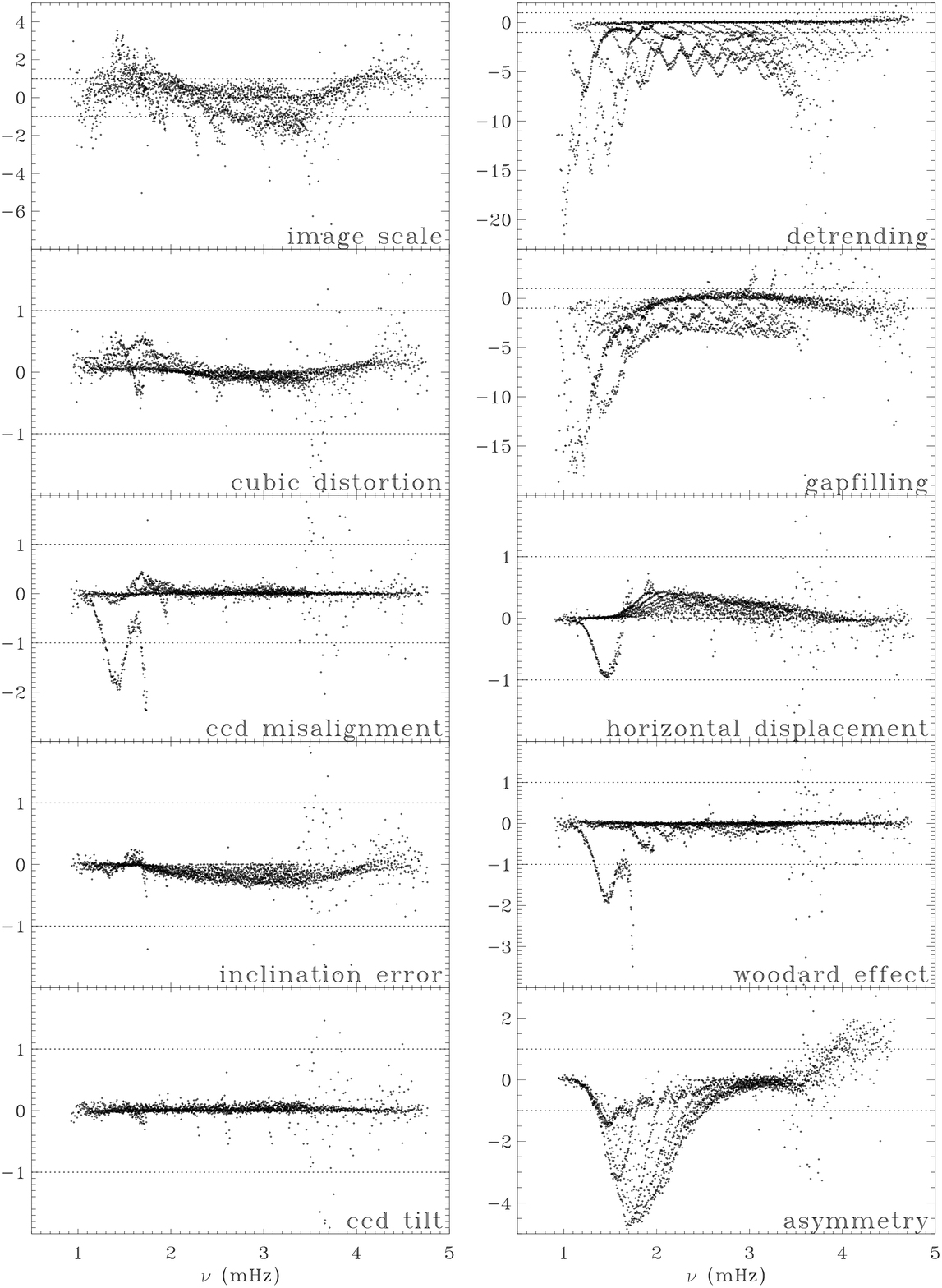}}

\caption{Change in background parameter resulting from each change as a
function of frequency, in units of standard deviation. Each panel is scaled differently;
dotted lines show the $\pm 1 \sigma$ levels.}

\label{F-bacdiff}
\end{figure}

\begin{figure}

\centering
{$\Delta a_1/\sigma_1$}\par\medskip

\centerline{\includegraphics[width=1.0\textwidth,clip=]{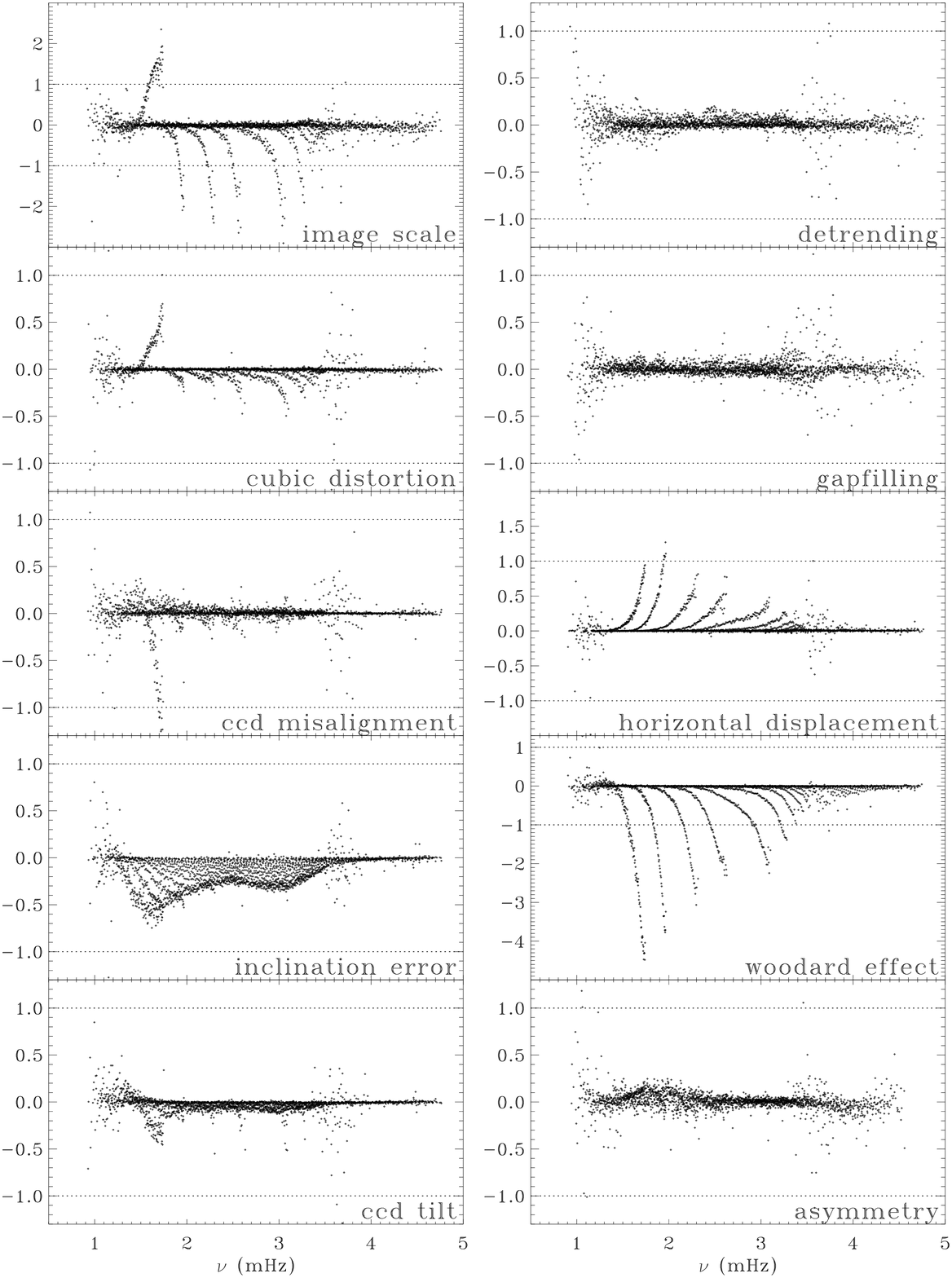}}

\caption{Change in $a_1$ resulting from each change as a function of
frequency, in units of standard deviation. Each panel is scaled differently;
dotted lines show the $\pm 1 \sigma$ levels.}

\label{F-a1diff}
\end{figure}

As can be seen in Figure \ref{F-nudiff}, the change in frequency was most
significant for the image-scale correction and asymmetric fits. Including
the horizontal displacement and correcting for distortion of
eigenfunctions made the next most significant changes, followed by
correcting for cubic distortion, in agreement with our previous work
\cite{improvements}.  Differences in detail between these and our previous
results can mostly be attributed to the different method that we have used for
computing mode averages; by first taking the common modeset for each 72
day interval, the calculation of the averages becomes much more
straightforward.  For the image-scale correction, some of the difference
in magnitude of the change in mode frequency can be attributed to the
different epoch we reanalyzed. Previously we studied the two years
beginning in January 2003, whereas in this work we study the three years
beginning in January 2004, and the image-scale error is the only problem
with the original analysis that is known to become worse over time.  For
the asymmetric fits, we used an improved iteration scheme for the
asymmetry parameter, which seems to have resulted in a smaller change in
frequency. The correction for CCD misalignment made a significant difference for the
$f$-mode, but otherwise this correction, the correction for the 
inclination error, the correction for CCD tilt, 
improved detrending, and improved gapfilling typically resulted in less than $0.5\,\sigma$ change in the mode 
frequencies.  We have also used a different
method for calculating the Woodard effect, as described above, but we
found this made less than a $0.5\,\sigma$ difference in all of the parameters
for the vast majority of modes.  Therefore in all plots we show only the
results of using the second method.

We find similar results for the amplitude and width (Figures
\ref{F-ampdiff} and \ref{F-widdiff}), although for both of these
parameters the detrending and gapfilling made much more significant
differences.  This is likely because these two changes in the processing
made the dominant changes to the background parameter (Figure
\ref{F-bacdiff}), as one might expect.  We also point out that the large
scatter of all three of these parameters just above 3.5\,mHz indicates an
instability of the fits in this frequency range, which may perhaps relate
to the bump as well.

The changes in $a_1$ (Figure \ref{F-a1diff}) have relative magnitudes that are 
roughly similar to the changes in frequency, the most notable exception
being that correcting for the Woodard effect caused the dominant changes
to this parameter.  For the $f$-mode, correcting for the image scale,
cubic distortion, and misalignment of the CCD resulted in changes with the same sign
as the frequency changes, but for the $p$-modes, and all modes when
correcting for horizontal displacement and the Woodard effect, the changes
had opposite sign.  The changes in $a_1$ resulting from the 
inclination correction were more significant than the frequency changes,
and show an interesting frequency dependence not seen in other parameters
for this correction.  The effects of the various changes on inversions of
$a_1$ are discussed in the next section.

\begin{figure}[h]
\centerline{\includegraphics[width=1.0\textwidth,clip=]{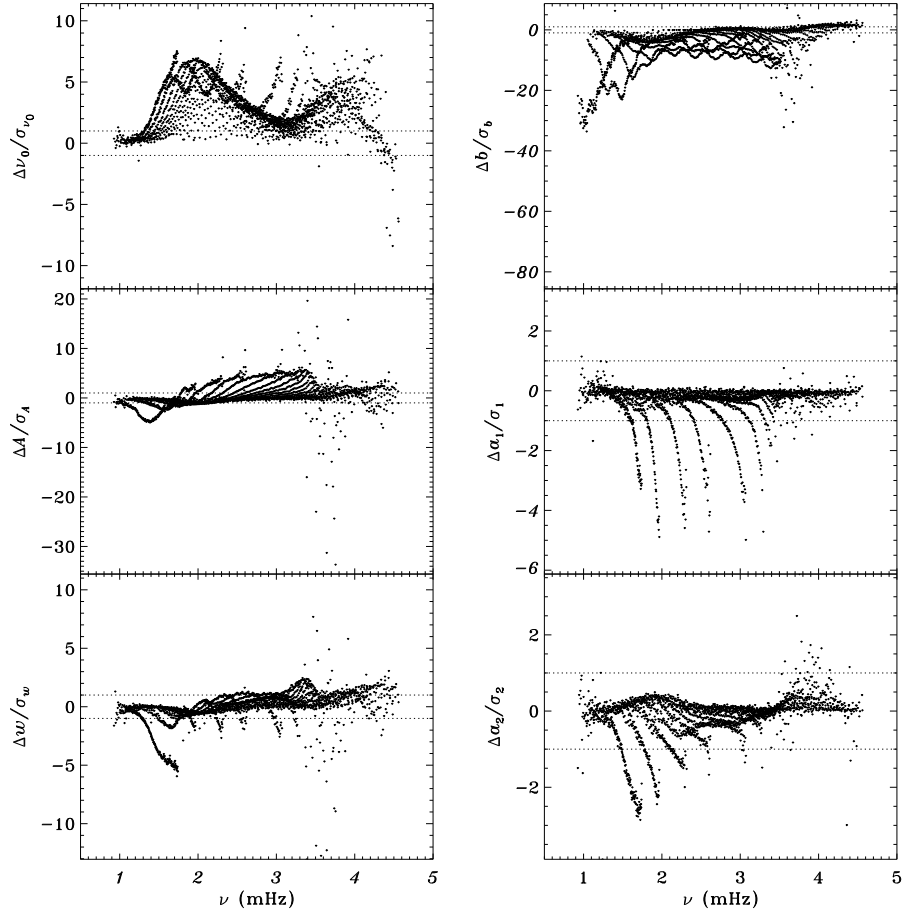}}

\caption{Change in frequency, amplitude, width, background parameter,
$a_1$, and $a_2$ resulting from all changes as a function of frequency, in
units of standard deviation, averaged over the entire mission.
Dotted lines show the $\pm 1 \sigma$ levels.
The range of these plots includes all points.}

\label{F-finit}
\end{figure}

\begin{figure}[h]
\centerline{\includegraphics[width=1.0\textwidth,clip=]{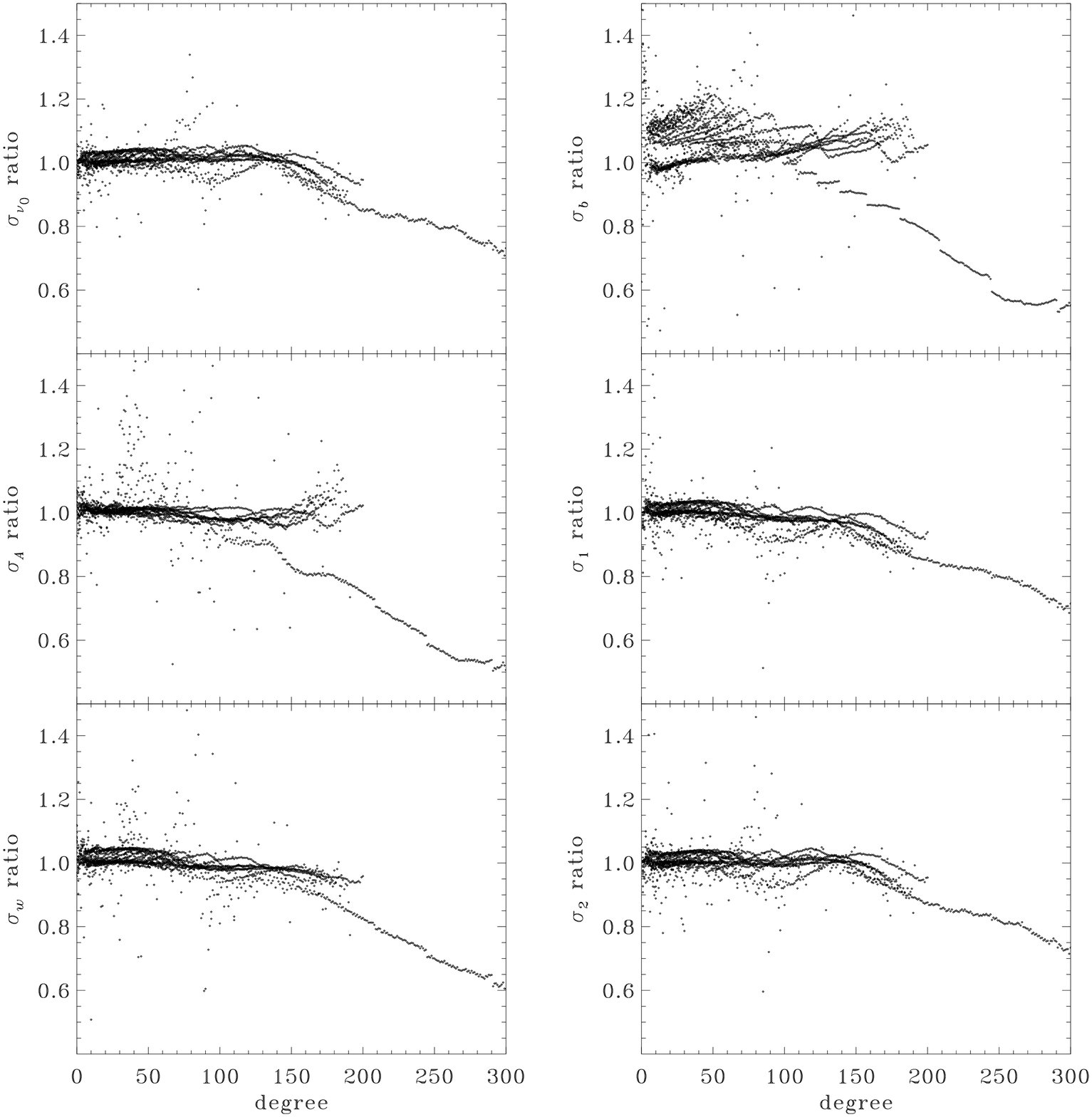}}

\caption{Ratio of the improved error estimates to the original error
estimates as a function of spherical harmonic degree for the parameters shown in Figure
\ref{F-finit}.  The improved estimates do not include fitting asymmetric
profiles.  For the background, 2.1\,\% of points do not fall within the
range shown on the plots.  For the other parameters, at most 0.6\,\% of
points are not shown.}

\label{F-finiterr}
\end{figure}

\begin{figure}[h]
\centerline{\includegraphics[width=1.0\textwidth,clip=]{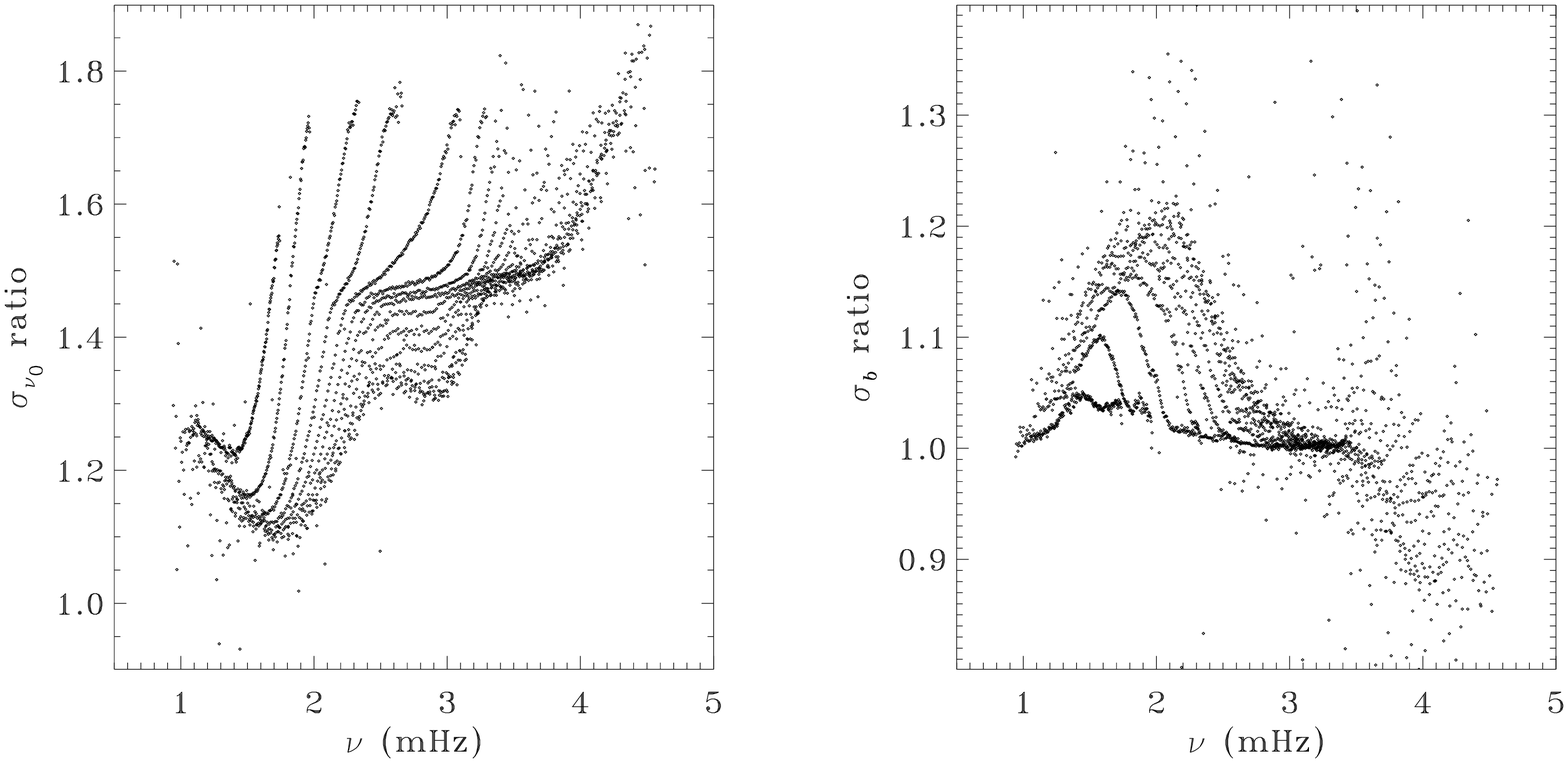}}

\caption{Amplification of errors for frequency and background resulting
from the use of asymmetric profiles.  For the background, 2.9\,\% of points
are excluded; for the frequency, 0.3\,\% are excluded.}

\label{F-asymerr}
\end{figure}

To see the effect of all of the changes in the processing taken together, we
examine the mission averages, formed as described above. Figure
\ref{F-finit} shows the result for various mode parameters.  For the
$p$-modes, the error estimates were mostly unaffected. However, the set of
all improvements up to and including the correction for the Woodard effect
resulted in substantially lower error estimates for the $f$-modes, as
shown in Figure \ref{F-finiterr}.  Unfortunately, using asymmetric line
profiles resulted in substantially higher error estimates for the mode
frequencies and background parameters, as shown in Figure \ref{F-asymerr}.

\begin{figure}[h!]
\centerline{\includegraphics[width=1.0\textwidth,clip=]{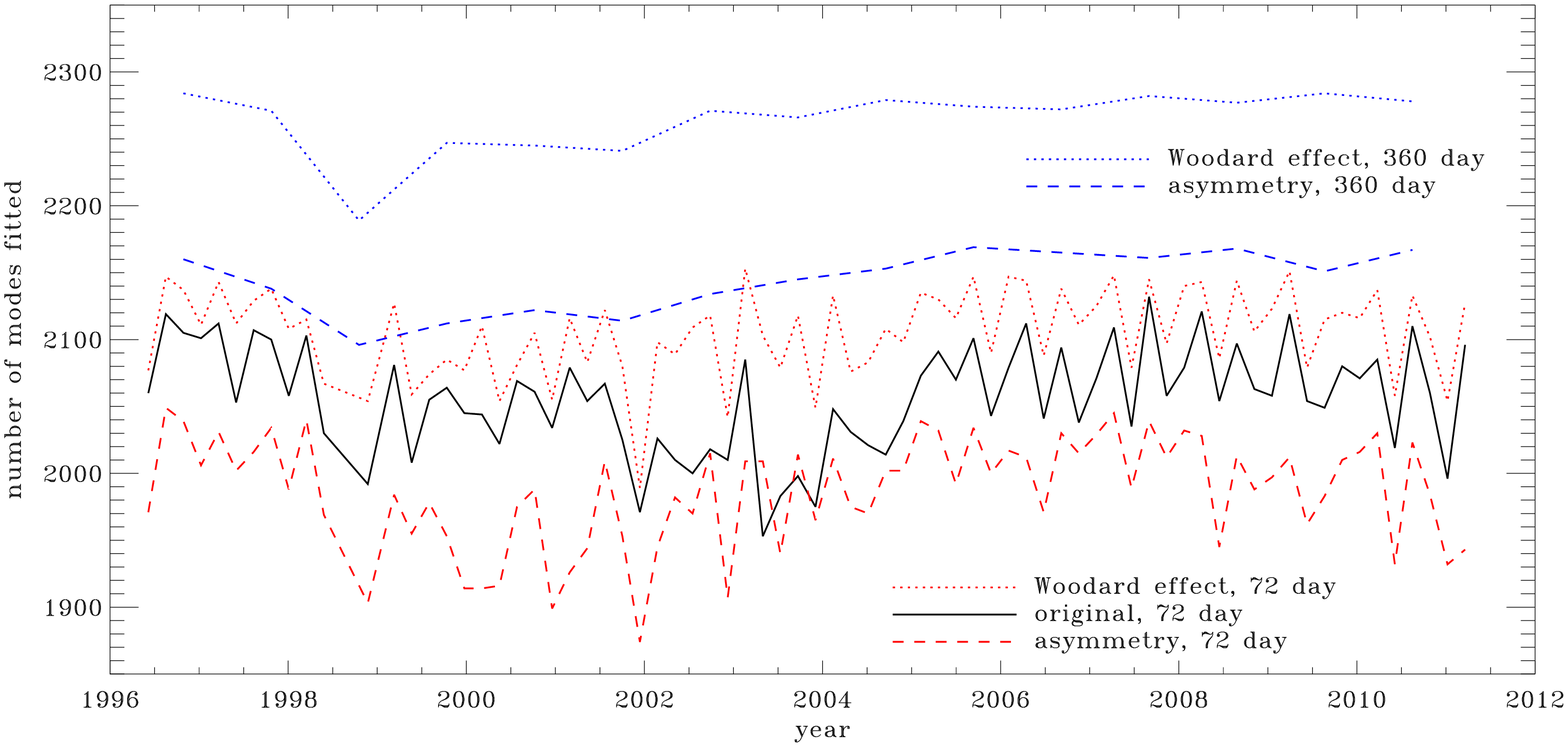}}

\caption{Number of modes fitted as a function of time for the five
different ways we analyzed the entire mission.  Dotted lines show the set
of all changes in the processing up to correcting for the Woodard effect;
dashed lines show the result of also using asymmetric line profiles.  In
both cases the higher line is for the 360-day fits, the lower line is for
the 72-day fits. The solid line shows the original analysis.}

\label{F-coverage}
\end{figure}

\begin{figure}[h!]
\centerline{\includegraphics[width=1.0\textwidth,clip=]{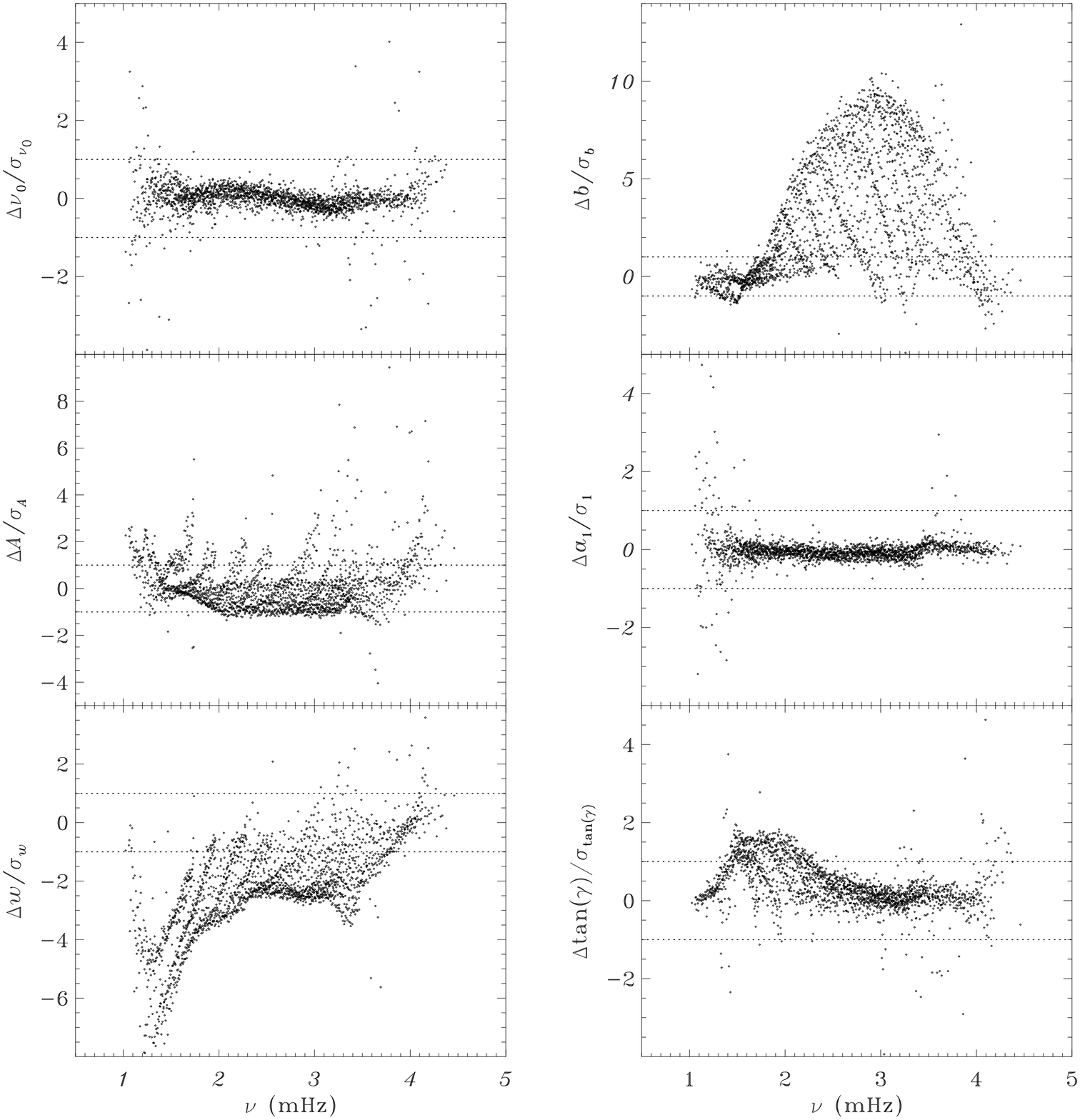}}

\caption{Difference in frequency, amplitude, width, background parameter,
$a_1$, and asymmetry parameter between 360-day fits and an average of 72-day 
fits as a function of frequency, in units of standard deviation from
the 360-day fits. Dotted lines show the $\pm 1 \sigma$ levels. 
The sense of subtraction is 360 day minus 72 day. At
most 0.9\,\% of points have been excluded.}

\label{F-asym360d}
\end{figure}

\begin{figure}[h!]
\centerline{\includegraphics[width=1.0\textwidth,clip=]{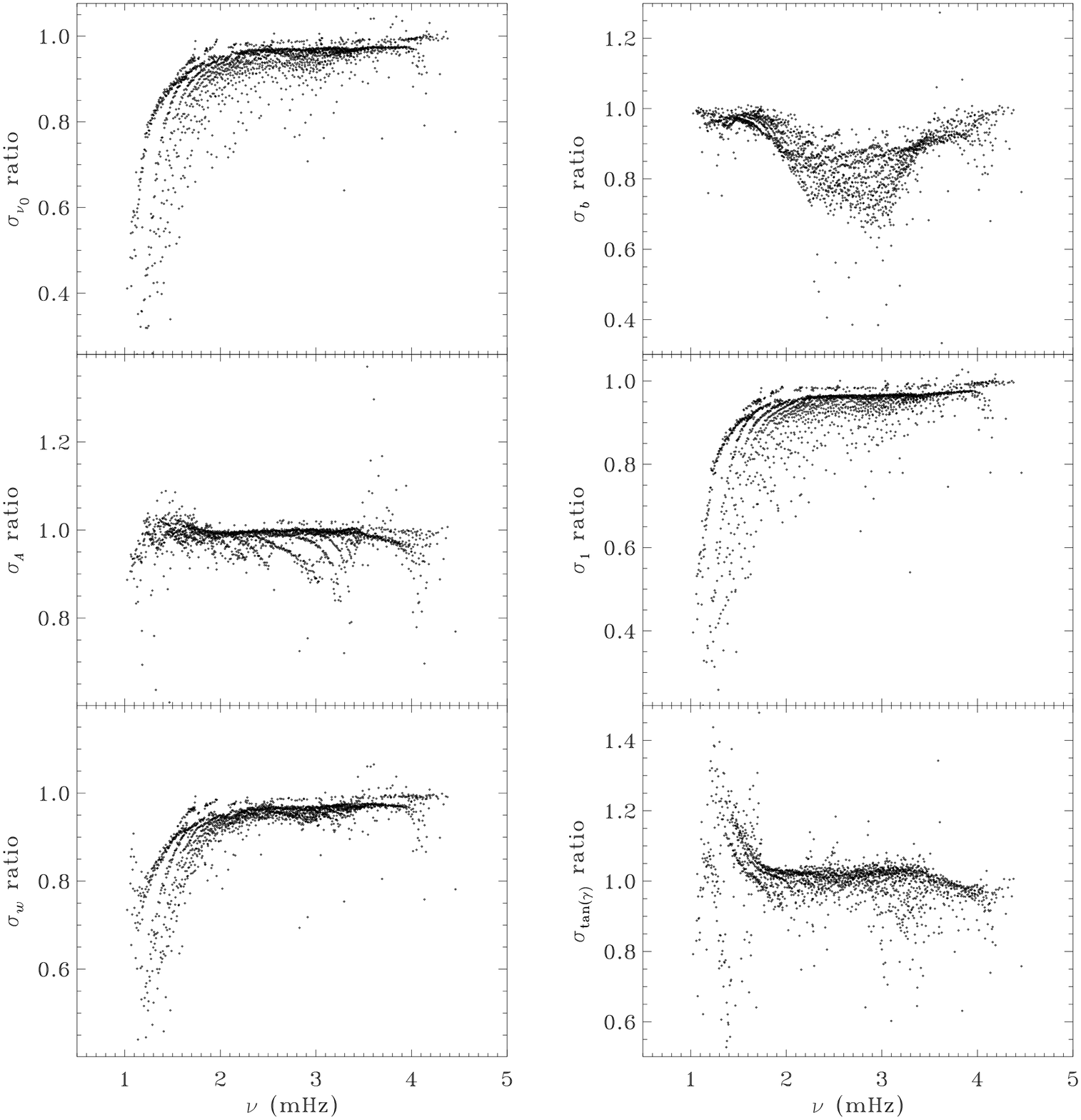}}

\caption{Ratio of the errors estimated from 360-day fits to the errors
estimated from an average of 72-day fits as a function of frequency for
the parameters shown in Figure \ref{F-asym360d}. At most 1.9\,\% of points
have been excluded.}

\label{F-asym360derr}
\end{figure}

One easy check of the robustness of our results is to compare the
72-day and 360-day analyses.  Even without examining any mode parameters,
one can see that the 360-day analysis was more successful in the sense
that it was able to fit more modes, as shown in Figure \ref{F-coverage}.  
To compare the mode parameters, for each 360-day interval we averaged 
the results of the five corresponding 72-day
analyses (three for the third 360-day interval) for the modes that were present in all
of them.  The errors used are the errors on the average.  Then we formed 
modesets common between the 360-day and 72-day analyses as above, representing the mission averages,  
this time taking the average error.  
The differences in mode parameters using asymmetric line profiles are
shown in Figure \ref{F-asym360d} and the corresponding error ratios are
shown in Figure \ref{F-asym360derr}.  The results were mostly similar using
symmetric line profiles.  To compare the background parameters, we
subtracted log(5) from the 360-day fits.

Although the change in frequency seems to show a weak systematic
dependence on frequency, the changes are mostly not significant. The
change in frequency was slightly more significant using symmetric line
profiles, especially at low frequencies. The changes in amplitude show
ridge structure; although the majority of modes show reduced amplitude,
the mean change is actually positive. The changes in width show ridges as
well, but here the width is almost always less for the 360-day fits, and
more so at lower frequencies. This is as one might expect, since the lorentzian is not well-resolved 
when the width is on the order of the width of a frequency bin. 
The increased frequency resolution of the 360-day fits better characterizes these low widths.  
The background parameter shows the most significant changes (an increase
except for the $f$-mode), but centered on the $p$-mode band, where the
noise is drowned by the signal.  The changes in $a_1$ are the flattest,
although a feature is discernible around 3.5\,mHz.  The asymmetry parameter
was in general greater for the 360-day fits, with a peak around 1.8~mHz.  
For the frequency, width, and $a_1$, the estimated errors were much lower
for the 360-day fits at low frequencies, again as one might expect.  
Harder to understand is why the error on the asymmetry parameter increased
in the same frequency range.  The background parameter also had lower
errors, but again in the center of the frequency range.

\subsection{Systematic Errors}
   \label{S-syserrs}

In this section we will refer to the changes in processing by the order 
in which they were applied.  This is summarized in Table \ref{T-corrlist}.

\begin{table}[h]

\caption{Sequence of changes made to the analysis; each analysis includes 
the changes made in all previous ones.}

\label{T-corrlist}
\begin{tabular}{ll}     
\hline
0 & original analysis \\
1 & image scale  \\
2 & cubic distortion \\
3 & CCD misalignment \\
4 & inclination error \\
5 & CCD tilt \\
6 & window functions and detrending \\
7 & gapfilling \\
8 & horizontal displacement \\
9 & distortion of eigenfunctions (``Woodard effect'') \\
10 & asymmetric line profiles \\
\end{tabular}
\end{table}

\begin{SCfigure}[][hb!]
\centering
\caption{Tradeoff curves for several analyses.  Dotted curve is for
original analysis.  Dash--dot curve shows first correction. Short-dashed
curve shows first three corrections.  Long-dashed curve shows first eight
corrections (note this curve is above the one for only the first
correction).  Solid line is for all corrections.  Symbols, from left to
right, indicate tradeoff parameters of $\mu=10^{-4}$ and
$\mu=10^{-9}$. }

\includegraphics[width=0.7\textwidth,clip=]{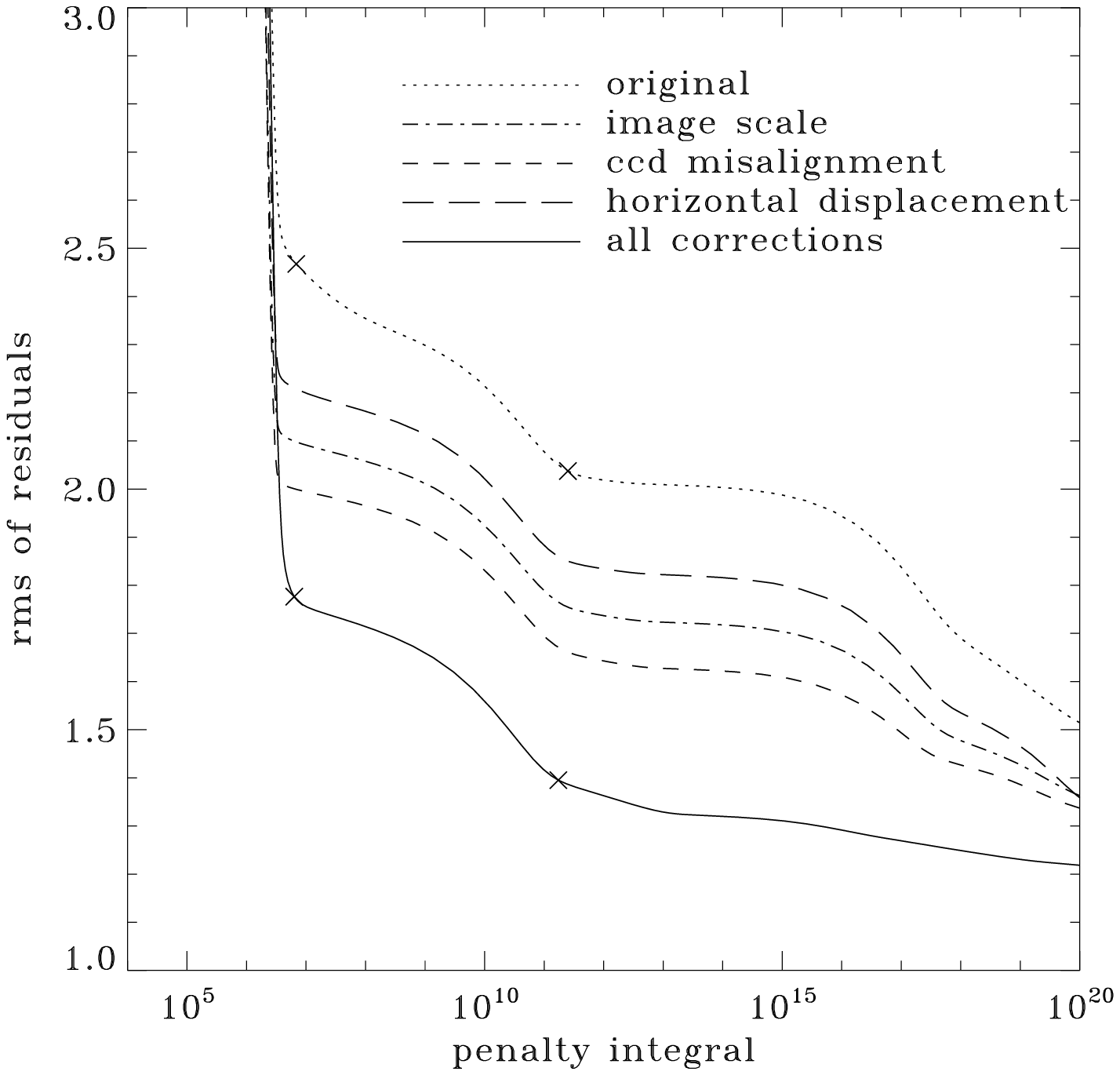}
\label{F-tradeoff1}
\end{SCfigure}

To see the effect of the various changes on our systematic errors, we
begin by performing simple one-dimensional regularized least-squares rotational 
inversions of the $a_1$-coefficient only.  An RLS inversion seeks to
minimize the sum of normalized residuals squared plus a penalty term
that serves to constrain rapid variations in the solution.  In particular,
we have chosen to minimize
\begin{equation} \label{E-rls}
\sum_{n\ell}\left[\frac{1}{\sigma_1(n,\ell)} 
\left( \int_0^1 K_{n\ell}(r)\bar{\Omega}(r){\rm d}r - a_1(n,\ell) \right)\right]^2 + 
\mu \int_0^1 \left(\deriv{^2\bar{\Omega}}{r^2}\right)^2 {\rm d}r
\end{equation}
where $\bar{\Omega}$ is the inferred rotation rate, the $K_{n\ell}$ are known
kernels calculated from the mode eigenfunctions that relate the rotation
rate to $a_1$, $\sigma_1$ is the standard error on $a_1$, $r$ is fractional
radius, and $\mu$ is the tradeoff parameter that controls the relative
importance of the two terms.  A low value of $\mu$ will fit the data
better, but the solution may oscillate wildly as a function of radius.  A
higher value of $\mu$ will attenuate this feature (the solution will be more
regularized)  at the cost of increased residuals (\opencite{schou94inv}).  
To choose a value of $\mu$, we have examined tradeoff curves, which are
constructed by varying $\mu$ and plotting the rms of the residuals against
the magnitude of the integral in the penalty term.  The changes in $a_1$ that underlie the
difference in the tradeoff curves for the different analyses were shown in
Figure \ref{F-a1diff}.  The tradeoff curves themselves (shown in Figure
\ref{F-tradeoff1}) were computed using a modeset constructed by finding
the modes common to all eleven analyses for each time interval and taking
the average in time over whatever modes were present; in this case the
errors used are the errors on the average.

As one can see, the image-scale correction made a substantial difference
to the tradeoff curve.  The curve for the cubic distortion correction is nearly
indistinguishable.  The correction for CCD misalignment made another significant
reduction in the residuals, but the curves for the next four changes to
the analysis all lie between the previous two.  Accounting for the
horizontal displacement caused a substantial increase in the residuals,
but accounting for the Woodard effect resulted in the lowest curve shown.  
The use of asymmetric profiles made no change to the tradeoff curve.  
This is basically in line with what one might expect based on the differences in 
$a_1$ resulting from each change in the analysis shown in
 Figure \ref{F-a1diff}.  

To choose a value of $\mu$, one typically looks for the
``elbow'' in the tradeoff curve: the place where the residuals stop
decreasing sharply, so that further decreases of $\mu$ will be of little
benefit.  Unfortunately, there seem to be two elbows in the
curves shown in Figure \ref{F-tradeoff1}.  For the initial and final
analyses, we have marked the point corresponding to the highest reasonable
value of $\mu$ ($10^{-4}$) and the lowest value one might reasonably use
($10^{-9}$). Furthermore, if the model were a good fit to the data, 
for the lowest values of $\mu$ the tradeoff curve should approach a value of 1.0,
which it does not.

\begin{figure}[h]
\centerline{\includegraphics[width=1.0\textwidth,clip=]{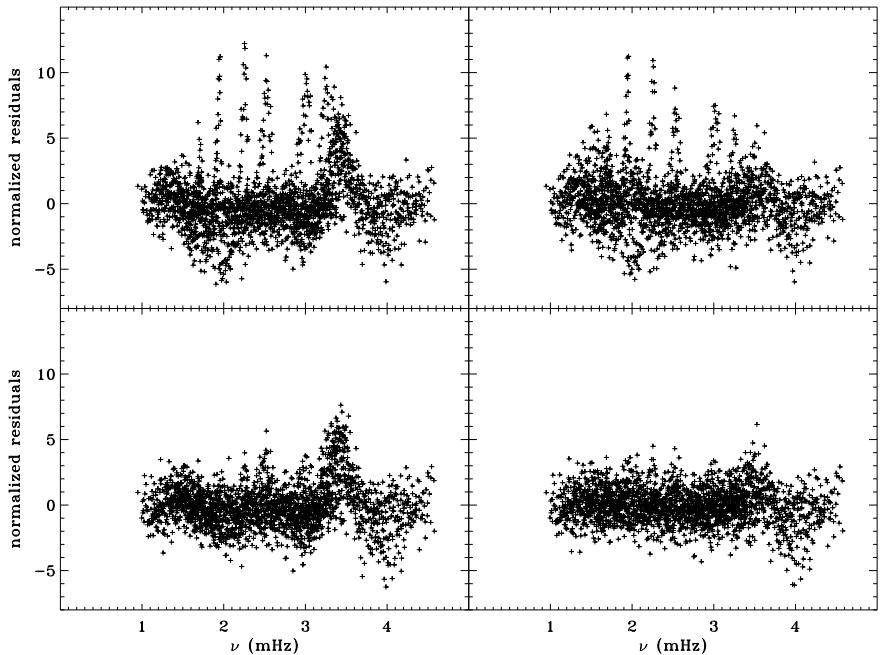}}

\caption{Normalized residuals as a function of frequency. Top panels show
original analysis, bottom panels show analysis with all changes applied.
Left panels show $\mu=10^{-4}$, right panels show $\mu=10^{-9}$.
The sense of subtraction is the opposite of Figure \ref{F-bumporig}
for ease of visual comparison.}

\label{F-bumps}
\end{figure}

\begin{SCfigure}[][h]
\centering

\caption{Internal rotation as a function of radius for the final analysis;
curves for original analysis are similar.  The solid line is the inversion
result using $\mu=10^{-4}$; the dashed line uses $\mu=10^{-9}$. }

\includegraphics[width=0.7\textwidth,clip=]{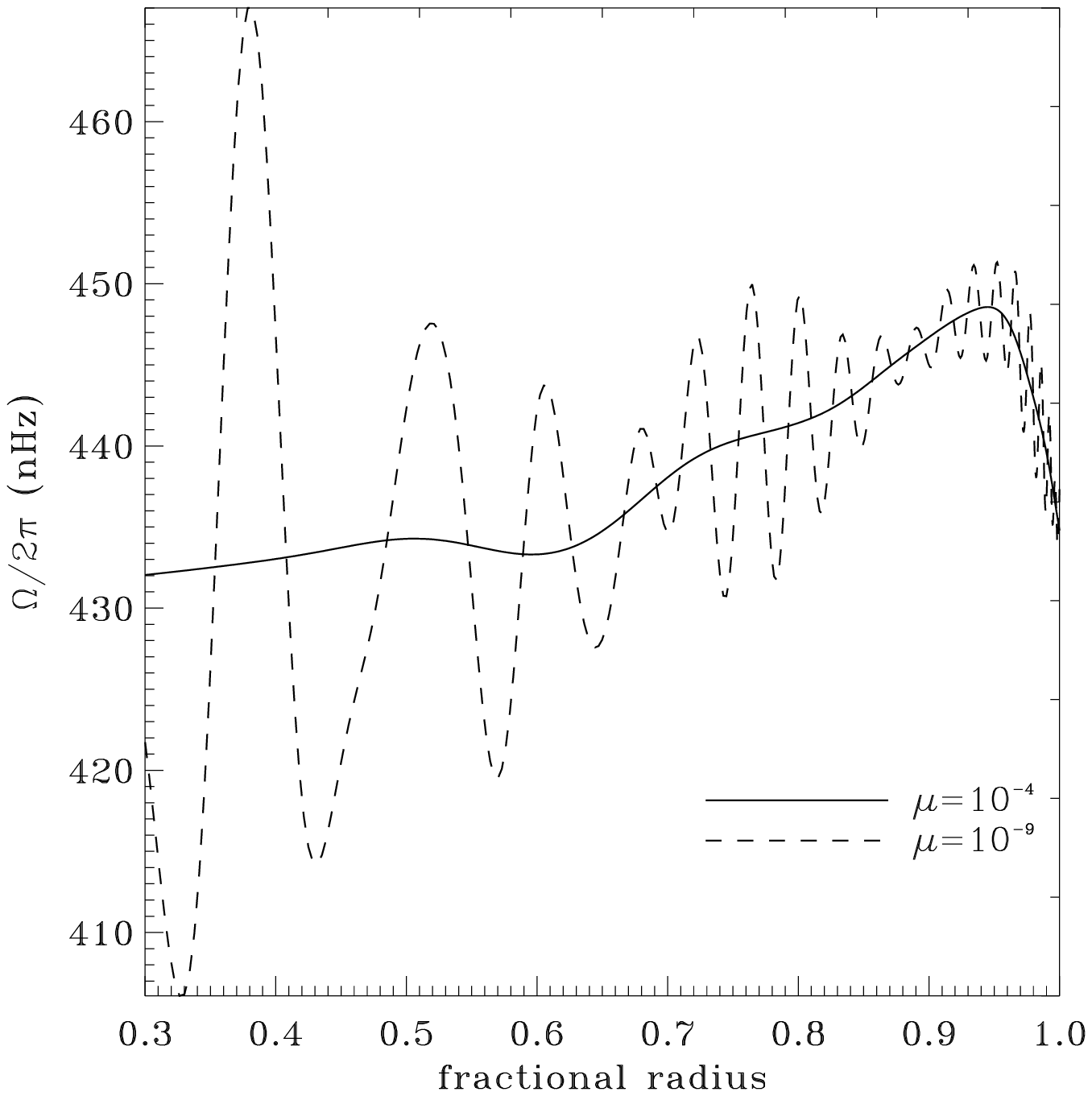}
\label{F-rotprof}
\end{SCfigure}

In Figure \ref{F-bumps} we show the normalized residuals of the inversions
for the original and final analyses and for the smallest and largest
values of $\mu$ given above. As one can see, the bump was mostly
unaffected by all the changes in the analysis.  A smaller value of $\mu$
decreases the size of the bump, but as Figure \ref{F-rotprof} shows, the
resulting rotation profile is unrealistic.  The fact that the
bump is only marginally present in the residuals for $\mu=10^{-9}$
suggests that this systematic error is responsible for the ``knee'' in the
tradeoff curves.  Notably, even this small value of $\mu$ was not able to
fit the horns in the original analysis, which are greatly reduced in
the final analysis.  This is likely the cause of the overall reduction in 
$\chi^2$.

To investigate the annual periodicity in the $f$-mode frequency
variations, we used the common modesets described above 
to fit a function of the form 
\begin{equation}
f(t) = A \sin(\omega_{\rm yr}t) + B \cos(\omega_{\rm yr}t) + Ct + D
\end{equation}
to the average fractional $f$-mode
frequency shift relative to its average over time, where
$\omega_{\rm yr}=2\pi/365.25$ and $t$ is measured in days.  We did separate
averaging and fits for four different ranges in degree [$\ell$]: 101 to 150,
151 to 200, 201 to 250, and 251 to 300.  In each case, for each $\ell$ we
took the average over whatever intervals it was fit in. Then, for each
interval, we took the difference between each $\ell$ and the time average,
divided by the time average, and then averaged over the range in $\ell$.  We
performed a weighted least-squares fit to this data, 
which yielded values for the parameters $A$, $B$, $C$, $D$, and their corresponding errors.

\begin{figure}[h]
\centerline{\includegraphics[width=1.0\textwidth,clip=]{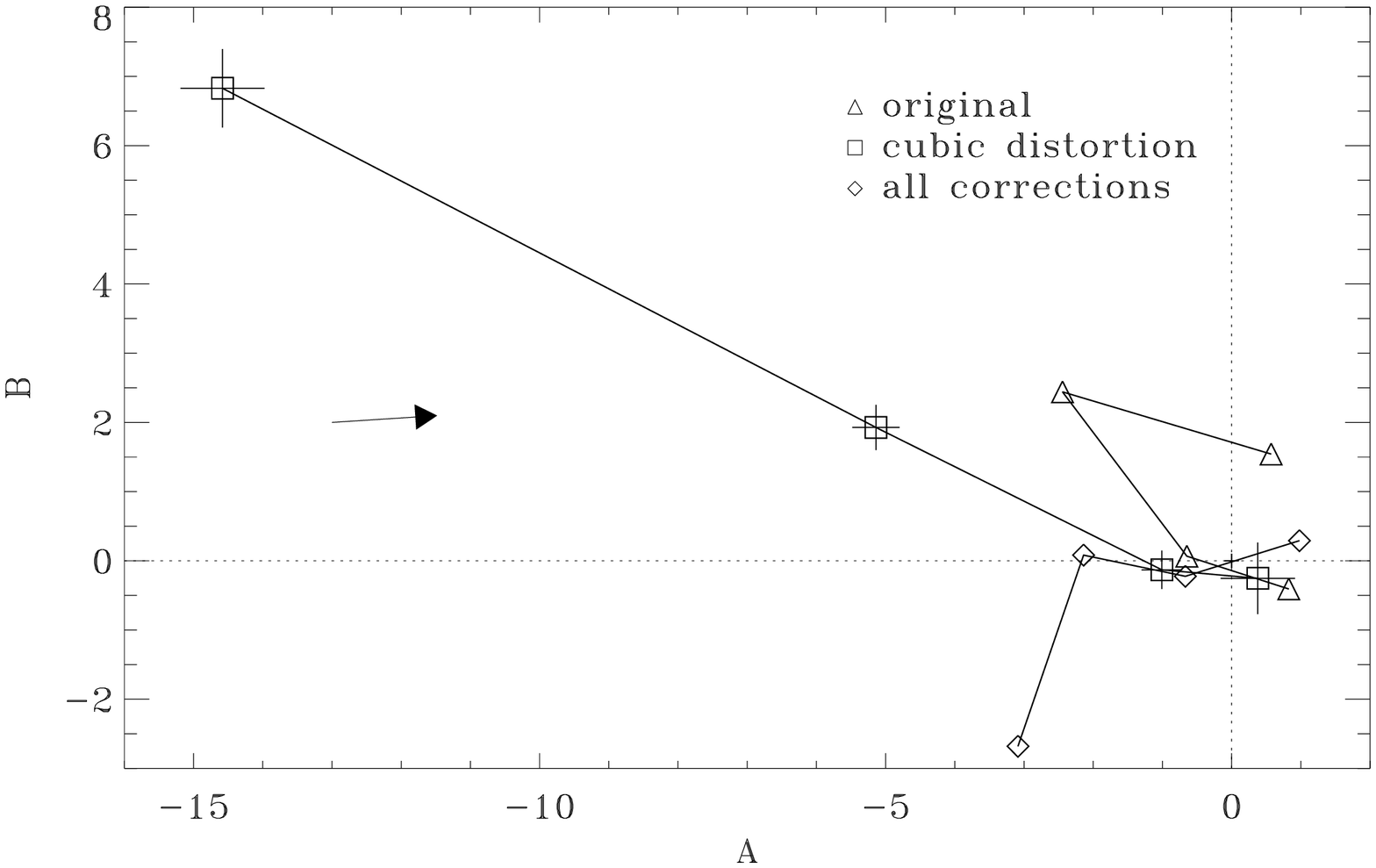}}

\caption{Amplitude of cosine vs. sine component of 
annual periodicity for three analyses after Doppler correction: 
triangles show original analysis, diamonds show final analysis, and squares show
the correction for cubic distortion, which yielded the largest amplitude of the 
annual component.  Solid lines connect points for different ranges in $\ell$, 
beginning with the lowest range on the lower right.  The arrow shows the size and direction of the 
shift resulting from the Doppler correction.  The errors on $A$ and $B$ were similar 
for all analyses; the error bars show an average value.  All values 
have been multiplied by $10^6$ to match the units in Figure \ref{F-radius}.}

\label{F-phase}
\end{figure}

The images produced by MDI, however, are taken at equal intervals of time
on the spacecraft, whereas it would be optimal if they were taken at equal intervals of time
on the Sun.  To correct for this effect, we applied the relativistic
Doppler shift due to the motion of the spacecraft.  That is, we multiplied
each frequency and its error by $\sqrt{(c+v)/(c-v)}$ where $c$ is the
speed of light and $v$ is the average velocity of the spacecraft away from 
the Sun, as derived from the \textsf{OBS\_VR} keyword of the input
dopplergrams for each 72-day interval.  The resulting fits are shown in
Figure \ref{F-phase}, as well as the shift caused by the Doppler
correction.

The amplitude of the annual component has a large variation between 
the different analyses, but in general it is always greater 
for the higher ranges in $\ell$.  The point in the plot for $\ell$=251\,--\,300 
of the original analysis contradicts that trend, but it 
must be noted that the fit represented by that point was an extremely 
poor one, which is likely related to the horns in the original 
analysis.  For the lower two ranges in $\ell$, the amplitude was only 
marginally significant.  Although not shown here, we note that 
the slope $C$ was zero for the lowest range in $\ell$, and becomes steadily 
more negative as $\ell$ increased, in agreement with 
previous findings \cite{antia01}.

\begin{figure}[h]
\centerline{\includegraphics[width=1.0\textwidth,clip=]{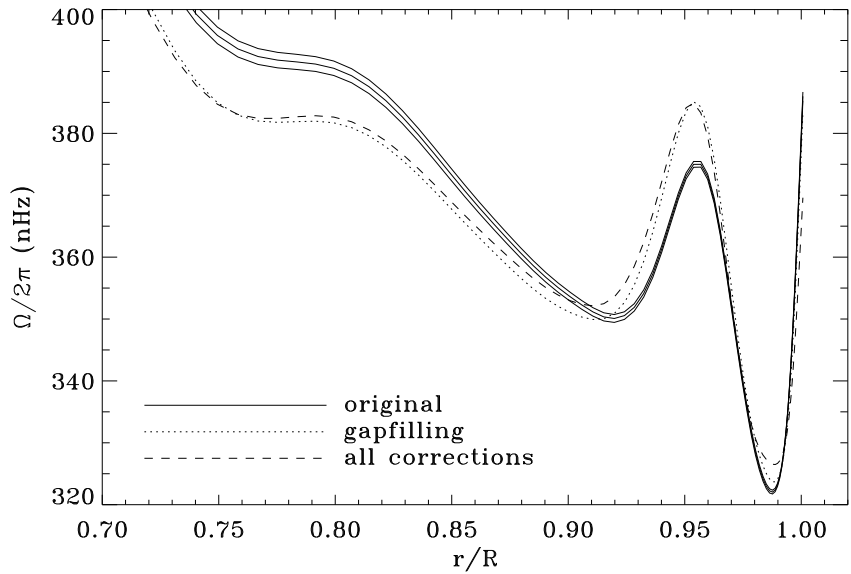}}

\caption{Internal rotation as a function of radius at $75^{\circ}$ latitude for three analyses.  Solid lines show the 
original analysis and its error bars; errors on the other analyses are similar. 
The dotted curve is the analysis that includes the improved gapfilling, and the 
dashed curve is the final analysis.}

\label{F-jet}
\end{figure}

Finally, to explore the anomalous peak in the near-surface rotation rate near the poles
(the high-latitude jet), we used the fits with 36 $a$-coefficients 
to perform two-dimensional RLS inversions 
for internal rotation. In this case we minimize 
\begin{eqnarray} \label{E-2drls}
\sum_{n\ell s}\left[\frac{1}{\sigma_{2s+1}(n,\ell)} 
\left( \int_0^1\int_0^\pi K_{n\ell s}(r,\theta)\bar{\Omega}(r,\theta){\rm d}r{\rm d}\theta - 
a_{2s+1}(n,\ell) \right)\right]^2 + \nonumber \\ 
\mu_r \int_0^1 \left(\deriv{^2\bar{\Omega}}{r^2}\right)^2 {\rm d}r + 
\mu_\theta \int_0^\pi \left(\deriv{^2\bar{\Omega}}{\theta^2}\right)^2 {\rm d}\theta
\end{eqnarray} 
in perfect analogy with Equation (\ref{E-rls}) \cite{schou94inv}.
We formed common modesets and averaged them using the same method described above for one-dimensional inversions, 
and used tradeoff parameters of $\mu_r=10^{-6}$ and $\mu_\theta=10^{-2}$ 
for the radial and latitudinal regularization terms respectively. 
Using this relatively high value for $\mu_\theta$ should dampen variations in latitude \cite{howe2000}. 
The results are shown in Figure \ref{F-jet};
the jet is more pronounced in this plot than in Figure \ref{F-rot2d}, 
which can be attributed both to the different modeset and 
to the smaller errors resulting from averaging.
Although in every updated analysis the polar jet actually had a greater magnitude than in the original analysis, 
the gapfilling resulted in a reduced rotation rate 
in the lower convection zone, which brings our result closer to agreement with inferences drawn by the GONG analysis
\cite{comparison1}.

\section{Discussion and Future Prospects}
     \label{S-discussion} 

We have found that the various changes that we made to the processing of
medium-$\ell$ data from MDI resulted in significant changes in mode
parameters.  In summary, changes in width were overall the least
significant, followed by the changes in $a_1$, which mostly resulted from
correcting for the distortion of eigenfunctions by the differential rotation (the Woodard effect).
 The background was largely unaffected by
most changes except the improved detrending and gapfilling.  The 
image-scale correction made the dominant changes to the amplitudes and
frequencies.  For the latter, large changes also resulted from accounting
for asymmetry, horizontal displacement, the Woodard effect, and cubic
distortion, in decreasing order of significance.

Not only is one led to believe these changes represent an improvement as a
matter of principle, but some of the systematic errors in the analysis
have been reduced as well.  In particular, the horns have been greatly
reduced, resulting in overall lower residuals from rotational inversions.  
A more stubborn systematic error is the bump in the odd
$a$-coefficients, which seems to be reflected in the anomalous shape of
the tradeoff curve.  This remained almost completely unchanged in all
analyses.  Nor did any change to the analysis make a reduction in the
high-latitude jet just below the solar surface, although there is an
improved agreement with GONG in the lower convection zone.

Regarding the annual periodicity in the $f$-mode frequencies, we found
that the first change that we applied, the image-scale correction, resulted in
a drastically increased magnitude of the annual component for the higher
two ranges in $\ell$.  The correction for cubic distortion resulted in an
even higher amplitude.  After correcting for the misalignment of the CCD, however,
the amplitude was reduced and did not vary much for later changes.  We
conjecture that the original fits were so poor at high $\ell$ (thus the horns)  
that the one-year period was swamped by noise there. 
The image-scale correction, which was the 
most significant one for the frequencies, itself has a one-year period 
due to its dependence on observer distance.  Hence this correction revealed 
the remaining annual periodicity in the $f$-mode frequencies, which appears to 
result mostly from errors in $P_{\rm eff}$.  Due to symmetry, one would expect the 
frequency error to depend on the absolute value of the error in $P_{\rm eff}$.  
The inclination error by itself would therefore be expected 
to result in a six-month period, but the combination with the misalignment 
of the CCD causes a one-year period.  Hence, although the correction 
for CCD misalignment is constant in time, it still greatly reduces the annual component.

Of concern to us is the discrepancy between the 360-day analysis, which in
principle should be more accurate, and the 72-day analysis.  Most notably,
it indicates a problem with our model of the background. Interestingly, the
asymmetry was the only parameter for which the error was greater for the
360-day fits (at low frequency), and adding the asymmetry also made significant changes to
the background and its error.

\begin{figure}[h]
\centerline{\includegraphics[width=1.0\textwidth,clip=]{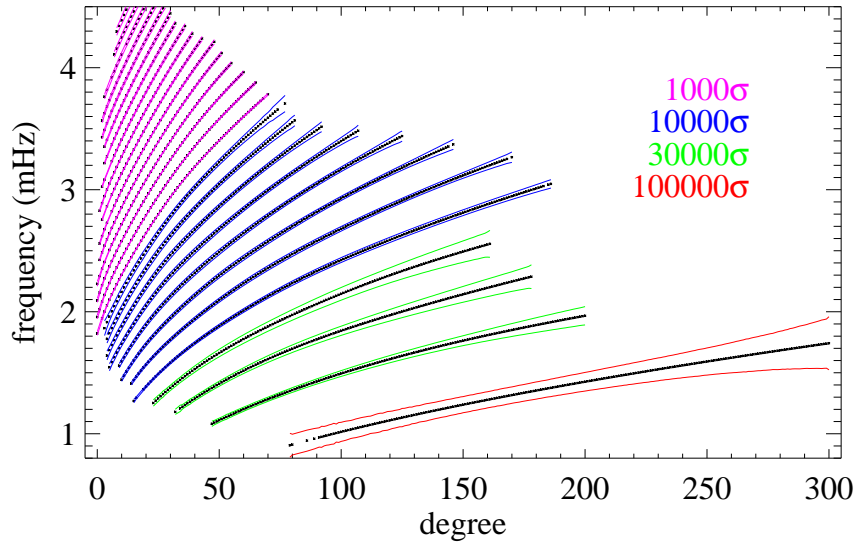}}

\caption{An $\ell$-$\nu$ diagram with magnified errors. Dots represent a mode that 
was fitted in at least 12 of the 15 years we
analyzed using symmetric profiles.  Solid lines show the errors: 
for the $f$-mode, these have been multiplied by 100\,000.  For $n=1,2,3$ 
the errors have been multiplied by 30\,000.  The next eight ridges ($n=4\,-\,11$) 
have errors multiplied by 10\,000.  The remaining
ridges have errors multiplied by 1000.}

\label{F-lnu}
\end{figure}

In spite of these shortcomings, the analysis of the MDI data in its 
entirety allows us to determine mode parameters with extraordinary 
precision.  This is illustrated in Figure \ref{F-lnu}, where we show mode 
coverage in the $\ell$--$\nu$ plane along with the estimated uncertainty on 
the frequencies.

Although our analysis has in general been very successful, the core 
peakbagging routines were written at a time when computational 
capabilities were far less than now.  A number of approximations which 
were necessary 20 years ago could now be lifted.  The current work is an 
attempt to remove some of these limitations.  Over the years, other 
workers in the field have also made contributions to the problem of 
inferring physical properties of the Sun from medium-$\ell$ MDI data. 
\inlinecite{vorontsov13} have proposed fitting power spectra 
for rotation directly, circumventing the need to measure frequencies. As an intermediate
step they have still done so, using more physically motivated spectral models and an 
analytically calculated leakage matrix. \inlinecite{korzennik05} has used sine multi-tapers
as power-spectrum estimators and fit widths and asymmetries as functions of $m$. 
\inlinecite{johan} have fit $m$-averaged spectra using a methodology 
that extends to high $\ell$.

A potential difficulty facing these efforts is the computation of the 
leakage matrix.  In general, the use of a leakage matrix 
should increase the stability of fits, but the results will then depend 
upon the assumptions that went into its calculation.  In particular, one might 
consider using leakage matrices calculated for different observer distances and 
values of $B_0$.  This has been done explicitly by \inlinecite{korzennik13} and 
analytically by \inlinecite{vorontsov13}. Others have attempted 
to fit the coupling of modes by subsurface flows, among them \inlinecite{schad13} and
\inlinecite{woodard13}.

Although a comparison between the results of these other investigators and our improved analysis
is still pending, all agree that some systematic errors remain in every analysis. 
These have been variously attributed to anisotropy in the point-spread function 
of MDI, failure to account for the height of formation in the solar atmosphere 
of the observed spectral line or the difference in light travel time between disk 
center and limb, and the effect of convective flows on the phase of the oscillations.

For us, there are a number of ways to move forward.  The most obvious is
the extension of this work to other datasets.  First and foremost of these
must be the MDI full-disk data, which will allow us to determine how
systematic errors and mode parameters might depend on the smoothing of the
medium-$\ell$ data and its apodization.  Because of its duty cycle the 
full-disk data cannot be used to study the annual periodicity in our results,
but now the {\it Helioseismic and Magnetic Imager} (HMI: \opencite{hmi}) onboard the {\it Solar
Dynamics Observatory} (SDO) has taken a long enough span of data for it to
be suitable for this purpose.  Phil Scherrer (private communication, 2014) has suggested that the one year
period may be related to the variable (in solar coordinates) width of the
gaussian used for smoothing the medium-$\ell$ data; an analysis of the MDI
medium-$\ell$ proxy from HMI should elucidate the issue.  Finally, a 
repetition of the comparison with GONG results is long overdue.  The 
original comparisons all used GONG classic data; now that GONG+ (\opencite{gongplus}) has been 
in place for over 13 years and software pipelines in both projects have 
been updated, the time has come to renew an investigation of the 
systematic differences between the two.

There still remain possibilities for progress with the MDI medium-$\ell$
dataset itself.  One that is suggested by the results of this article is to
correct the timeseries for the relative motions of SOHO and the Sun.  
Although we can correct the frequencies after the fitting by Doppler
shifting them, there is no obvious way to correct the other mode
parameters.  Another change in the analysis that suggests itself is to the
width of the fitting window, since this is one of the things most notably
different in the GONG analysis and is also known to affect the shape of
the bump in the $a$-coefficients.  During the remapping performed prior to
spherical harmonic decomposition, we could implement an interpolation
algorithm that takes into account the correlation between points introduced
by the gaussian smoothing.  We have also considered the common practice of
zero-padding our timeseries before performing Fourier transforms.  
Lastly, the parameter space of the detrending and gapfilling remains
almost entirely unexplored.

%%%%%%%%%%%%%%%%%%%%%%%%%%%%%%%%%%%%%%%%%%%%%%%%%%%%%%%%%%%%%%%%%%%%%%%%%%%
%% Acknowledgements

\begin{acks}

This work was supported by NASA Contract NAS5-02139. SOHO is a mission
of international cooperation between NASA and ESA. The authors thank
the Solar Oscillations Investigation team at Stanford University and its
successor, the Joint Science Operations Center. We thank Rasmus Larsen 
in particular for providing the gapfilling code. Much of the work presented 
here was done while J. Schou was 
at Stanford University. T.P. Larson thanks
the Max-Planck-Institut f\"{u}r Sonnensystemforschung for generously
hosting him during the composition of this article.
\\ \\
\noindent
{\bf Disclosure of Potential Conflicts of Interest}
The authors declare that they have no conflicts of interest.

\end{acks}

%%%%%%%%%%%%%%%%%%%%%%%%%%%%%%%%%%%%%%%%%%%%%%%%%%%%%%%%%%%%%%%%%%%%%%%%%%%
%% Appendix

\appendix
%\section*{Data Access}
%\label{S-appendix}

Detailed information on how to access MDI data from the global
helioseismology pipeline can be found on the website of the Joint Science Operations Center (JSOC) at
%\href{http://jsoc.stanford.edu/MDI/MDI_Global.html}{jsoc.stanford.edu/MDI/MDI\_Global.html}.  
\url{http://jsoc.stanford.edu/MDI/MDI_Global.html}.
This page contains
documentation describing how the datasets used in this article were made and
how they can be remade.  In this appendix we describe how to access the
relevant archived data. In what follows we assume some familiarity
with the Data Record Management System (DRMS), detailed documentation for
which is linked from the above website.

Mode parameter files (as ASCII tables) for every analysis discussed in this paper are
available in the electronic supplementary material.  For the original
analysis, they (and a helpful \textsf{Readme} file) can also be found at
%\href{http://sun.stanford.edu/~schou/anavw72z/}{sun.stanford. edu/\~schou/anavw72z/}.  
\url{http://sun.stanford.edu/~schou/anavw72z/}.
For all other analyses,
they can also be retrieved from JSOC. The fields of a mode-parameter file
are the following: $\ell$, $n$, $\nu_0$, $A$, $w$, $b$, $x$, $\{\tan(\gamma)\}$,
$\sigma_{\nu_0}$, $\sigma_A$, $\sigma_w$, $\sigma_b$, $\sigma_x$,
$\{\sigma_{\tan(\gamma)}\}$, $a_1$, $a_2$, ... $a_N$, $\sigma_1$, $\sigma_2$,
... $\sigma_N$.  The parameter $\tan(\gamma)$ and its error will not be present
for fits done with symmetric profiles.  The value of $N$ is either 6, 18,
or 36.  Any parameter with zero error has not been fit for. The parameter 
$x$ is not fit for in these analyses and is retained for historical reasons.

The data for the different ``corrections'' are labelled by the strings
\textsf{corr1} to \textsf{corr9} corresponding to the numbering scheme in Table
\ref{T-corrlist}.  The final correction in this set refers to the first
way of applying the Woodard effect (holding $B_1$ and $B_2$ constant).
These data have all been generated in the first author's name space, with
mode parameters found in \textsf{su\_tplarson.corr\_vw\_V\_sht\_modes}.  
The primekeys are \textsf{T\_START}, \textsf{LMIN}, \textsf{LMAX}, \textsf{NDT}, 
and \textsf{TAG,} where \textsf{T\_START} is the
beginning of the corresponding timeseries, most easily specified by the
MDI day number suffixed by ``d'' (see Table \ref{T-data}). For all records
in this series, \textsf{LMIN}=0, \textsf{LMAX}=300, and \textsf{NDT}=103680, so these primekeys need
never be specified.  The \textsf{TAG} keyword is the label string, so \textsf{TAG} and 
\textsf{T\_START} uniquely specify every record.

The second way of applying the Woodard effect, as well as the asymmetric
fits, are both represented in the official MDI name space (\textsf{mdi}).  For the
former, mode parameters can be found in \textsf{mdi.vw\_V\_sht\_modes} and 
for the latter in \textsf{mdi.vw\_V\_sht\_modes\_asym}.  The primekeys are the same as
given above, with the exception that these series do not have the \textsf{TAG}
keyword and that \textsf{NDT}=518400 for the 360-day fits.  In addition, the
results used in this article have the \textsf{VERSION} keyword (not a primekey) in
these series set to \textsf{version2}.  If these data are reprocessed in the
future, \textsf{VERSION} will get a new value, but old versions can easily be retrieved.

The dataseries containing timeseries and window functions in the \textsf{mdi} name
space have also been archived and can be retrieved; details on these data
products are given on the above website.  The corresponding data in the
\textsf{su\_tplarson} name space have not been archived, but can be recreated if
needed.  The procedure for doing so can be found on the website.
The original timeseries and window functions have been archived in the
\textsf{dsds} namespace, but have not yet been ported to the standard DRMS format
for global helioseismology data products.  They can, however, still be
retrieved by request.

%%% %%%%%%%%%%%%%%%%%%%%%%%%%%%%%%%%%%%%%%%%%%%%%%%%%%%%%%%%%%%
%% Bibliography
%
% Using BibTeX
%
%\bibliographystyle{spr-mp-sola}
% %\bibliographystyle{spr-mp-sola-cnd} %% Alternative style: no title, no concluding page
\bibliography{paper}

\end{article} 
\end{document}